\documentclass[twocolumn,showpacs,preprintnumbers,amsmath,amssymb,prb]{revtex4}

\usepackage{graphicx}
\usepackage{dcolumn}
\usepackage{bm}

\begin{document}


\title{
Imaginary-time formulation of steady-state nonequilibrium
in quantum dot models
}

\author{J. E. Han}
\affiliation{
Department of Physics, State University of New York at Buffalo, Buffalo, NY 14260, USA}

\date{\today}

\begin{abstract}
We examine the recently proposed imaginary-time formulation for strongly
correlated steady-state nonequilibrium for its range of validity and
discuss significant improvements in the analytic continuation of
the Matsubara voltage as well as the fermionic Matsubara frequency.
The discretization error in the
conventional Hirsch-Fye algorithm has been compensated in the Fourier
transformation with reliable small frequency behavior of self-energy.
Here we give detailed discussions for generalized spectral
representation ansatz by including high order vertex corrections
and its numerical analytic continuation
procedures. The differential conductance calculations agree accurately
with existing data from other nonequilibrium transport theories.
It is verified that, at finite source-drain voltage, the
Kondo resonance is destroyed at bias comparable to the Kondo
temperature. Calculated coefficients in the scaling relation of
the zero bias anomaly fall within the range of experimental estimates.
\end{abstract}


\maketitle

Progress in nanoscale fabrication techniques has recently
generated a great deal of interest in electron transport out of equilibrium.
Well-controlled quantum dot (QD) structures in semiconductor devices have
enabled thorough investigation of quantum many-body effects in confined
geometry. One of the established phenomena is the zero bias anomaly
(ZBA) due to the Kondo effect in Coulomb blockade
devices~\cite{cronenwett,vanderwiel,grobis}, where the scaling behavior of nonlinear
conductance has been extensively verified. Recently, more complex quantum
dot systems such as molecular nano-junctions~\cite{scott} and multi-channel Kondo
dots~\cite{potok} with intricate device design have fueled intense
research for electron transport mechanism in strongly correlated regime.
Nanoscale single-electron devices hold great
promise not only in applications for quantum devices but also in
development of general quantum many-body theory out of equilibrium.

In the past few years, the strong correlation community has embraced the
challenge of developing quantum many-body formulations out of
equilibrium, which has lead to significant advances at various levels of
theories. Unlike equilibrium quantum many-body theory where the analytic
and numerical theories play complementary roles, the nonequilibrium
theory has only recently had full-fledged numerical tools which could
support or disprove
the diagrammatic approximations, known as Keldysh
technique~\cite{rammer,datta_book}. So far, numerous
algorithms have been proposed. However, most of the theories have yet to
be fully established to have reliable predictive power in a wide range of
strongly correlated transport. 

\begin{figure}[bt]
\rotatebox{0}{\resizebox{3.1in}{!}{\includegraphics{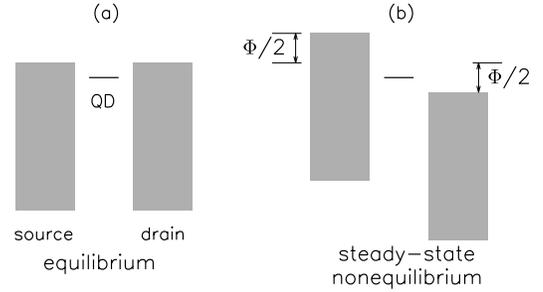}}}
\caption{(a) Source and drain reservoirs for electron and a quantum dot
level. Chemical potentials of the reservoirs are the same in
equilibrium. (b) Voltage bias $\Phi$ in steady-state
nonequilibrium divided between chemical potentials of the reservoirs. 
The quantum dot
energy level can be arbitrarily positioned with respect to the chemical potentials.
}
\label{fig:scheme}\end{figure}

The main focus of the theoretical efforts has been the description of
the transient behavior toward a nonequilibrium steady-state or the
properties of an established steady-state. Here the steady-state concerns the dc
current-carrying state driven by a time-independent bias in quantum dot
geometry, as sketched in Fig.~\ref{fig:scheme}. 
Quantum simulations of real-time behavior of nonequilibrium 
steady-state have been performed using time-dependent numerical 
renormalization group (tNRG~\cite{andersschiller,anders}), time-dependent density-matrix
RG (tDMRG~\cite{boulat,feiguin}), perturbative RG
(PRG~\cite{rosch,kehrein}), functional
renormalization group (fRG~\cite{jacobs,schmidt}), iterative summation
of real-time path-integral method (ISPI~\cite{eckel,weiss}),
diagrammatic Monte Carlo~\cite{schiro,werner}. 
Analytic methods of nonequilibrium Bethe ansatz~\cite{mehta} and
perturbative steady-state expansion~\cite{doyon} have been developed.

The imaginary-time formulation for steady-state nonequilibrium proposed
by Han and Heary~\cite{prl07} takes quite a different approach from the
above real-time techniques. Main motivation has been to extend the
equilibrium quantum many-body theory within a similar mathematical
framework of the imaginary-time formalism and therefore to easily adopt
existing equilibrium numerical techniques for complex quantum
interactions such as molecular dots~\cite{prb10}. The method combines
the equilibrium many-body theory and steady-state nonequilibrium quantum
statistics by extending the chemical potentials into complex variables,
the \textit{Matsubara voltage}. It has been shown that, upon analytic
continuation of complex chemical potential back to physical chemical
potential the theory recovers the nonequilibrium dynamics, and the
imaginary-time Green's functions map to real-time Green's functions.

In this review, we show how the previously introduced spectral ansatz
can be extended to include vertex corrections and give 
comprehensive discussions detailing the justification and the range of
validity of the analytic continuation. We further enhance the
computational method by modifying the conventional
Hirsch-Fye~\cite{hirsch} quantum Monte Carlo (QMC) algorithm by
compensating the discretization error and obtain electron self-energy with
much improved reliability. These improvements lead to reliable
differential conductance curves in accurate agreement with other
existing methods. We confirm that the Kondo resonance is
destroyed at the bias of Kondo scale and the resulting scaling behavior
of conductance is consistent with experiments.

The paper is organized as follows. In the following
section~\ref{sec:imag}, we define the imaginary-time Hamiltonian for
nonequilibrium. We then show the equivalence of imaginary-time and real-time
Green's functions in the perturbation expansion with arbitrary interaction
through the analytic continuation.
We give examples of different quantum dot geometry where
this formalism can be extended. In the following
section~\ref{sec:algorithm}, we briefly introduce the QMC procedure and
propose an algorithm which fixes discretization errors in the
nonequilibrium QMC. In section~\ref{sec:analytic} we extend the
previously introduced analytic continuation ansatz via Pad\'e
approximants and describe detailed numerical procedures to fit spectral
functions. We make a direct comparison of conductance in the small to
intermediate interaction regimes with other methods.
In section~\ref{sec:results} computational results are
presented and compare them with existing theories and experiments. In
the appendix, analytic structure of high order corrections to self-energy
is discussed.

\section{Imaginary-time formalism}
\label{sec:imag}

We start the discussion of nonequilibrium quantum theory with the
understanding that the one of the fundamental problems is that there are
two different operators governing the quantum statistics and the
time-evolution. In equilibrium, the time-evolution is given by
$e^{-itH}$ while the Boltzmann factor $e^{-\beta(H-\mu N)}$ provides the
quantum statistics with the chemical potential $\mu$ and the number
operator $N$. With a number-conserving Hamiltonian, $[H,N]=0$, the
Boltzmann factor and the time-evolution operator commute. For this
reason, the chemical potential $\mu$ is often set as the reference
energy ($\mu=0$) without losing generality. However in nonequilibrium of
quantum dot systems, bias voltage creates multiple electronic
chemical potentials in the source and drain reservoirs. Once the
tunneling between the quantum dot and the reservoirs is allowed and
many-body interactions are turned on, the reservoir states mix with the
time-evolution. Hence, the task of finding commutable nonequilibrium
density matrix and the time-evolution operators becomes a challenging
task. 

From the 1960s and 1970s, efforts have been made to formulate a Gibbsian
statistical mechanics in steady-state nonequilibrium using the Liouville
operator formalism~\cite{zubarev}. In 1993, Hershfield~\cite{hershfield}
has revisited the problem in the context
of the mesoscopic transport and has provided a formal proof in a compact
form of a density matrix for nonequilibrium steady-state. The idea
consists of decomposing the full Hamiltonian in terms of the formal
solution of scattering states of Lippman-Schwinger
equation~\cite{merzbacher}, and
applying different chemical potentials to scattering states derived from
each reservoirs. With the scattering state operator
$\psi^\dagger_{\alpha k\sigma}$ with the continuum index $k$, reservoir
index $\alpha=L,R$ (or $+,-$, respectively) and the spin index $\sigma$, the nonequilibrium
density matrix operator at the voltage bias $\Phi$ is written as
\begin{equation}
\hat{\rho}=e^{-\beta(\hat{H}-\Phi\hat{Y})},
\end{equation}
with the operator $\hat{Y}$
\begin{equation}
\hat{Y}=\frac12\sum_{k\sigma}(\psi^\dagger_{L k\sigma}\psi_{L k\sigma}
-\psi^\dagger_{R k\sigma}\psi_{R k\sigma}).
\end{equation}
However, despite its conceptual breakthroughs, Hershfield's idea has not
found practical implementations since fully interacting scattering
states cannot be known \textit{a priori}. The method has been adapted only in the
non-interacting models and in perturbative limits~\cite{prb06,prb07}.

To overcome such difficulties, Han and Heary have proposed an
imaginary-time formalism which constructs a formal perturbation
expansion from a non-interacting nonequilibrium steady-state.  The main
advantage of the method over the original Hershfield's idea is that
there is no need to construct the Hershfield's $\hat{Y}$-operator in the
interacting limit.
Many-body interactions are considered rigorously in a perturbation
series at arbitrary
order built on the non-interacting nonequilibrium density matrix
$e^{-\beta(H_0-\Phi Y_0)}$ with the $\hat{Y}_0$-operator computed
without many-body interactions.
To compensate the discrepancy between the time-evolution and the
Boltzmann factor, we introduce a mathematical trick of the Matsubara voltage.

Electron tunneling of single quantum dot connected to source and drain reservoirs
is modeled by 
\begin{eqnarray}
\hat{H}_0&=&\hat{H}_{0c}+\hat{H}_{0d}+\hat{H}_{0t}\\
\hat{H}_{0c}&=&\sum_{\alpha k\sigma}\epsilon_{\alpha k}c^\dagger_{\alpha
k\sigma}c_{\alpha k\sigma}\\
\hat{H}_{0d}&=&\epsilon_d\sum_\sigma d^\dagger_\sigma
d_\sigma\\
\hat{H}_{0t}&=& -\sum_{\alpha
k\sigma}\frac{t_\alpha}{\sqrt\Omega}(d^\dagger_\sigma
c_{\alpha k\sigma}+h.c.),
\end{eqnarray}
with $\hat{H}_{0c},\hat{H}_{0d},\hat{H}_{0t}$ for decoupled reservoirs,
quantum dot states, and QD-reservoir tunneling, respectively.
$c^\dagger_{\alpha k\sigma}$ are the fermion operators for 
continuum state,
$d^\dagger_\sigma$ quantum dot orbital operator, $t_\alpha$ the tunneling parameter
and $\Omega$ the volume of the reservoirs. Then the
non-interacting scattering state can be readily written down
as~\cite{prb06,prb07}
\begin{eqnarray}
\psi^\dagger_{0, \alpha k\sigma}
&=&c^\dagger_{\alpha
k\sigma}-\frac{t_\alpha}{\sqrt\Omega}g^0_d(\epsilon_{\alpha
k})d^\dagger_\sigma \nonumber \\
&&+\sum_{\alpha' k'}\frac{t_\alpha
t_{\alpha'}}{\Omega}\frac{g^0_d(\epsilon_{\alpha k})}{\epsilon_{\alpha k}
-\epsilon_{\alpha' k'}+i\eta}c^\dagger_{\alpha' k'\sigma}
\label{eq:psi}
\end{eqnarray}
with the non-interacting retarded Green's function
$g^0(\omega)=[\omega-\epsilon_d+i\Gamma]^{-1}$. $\Gamma$ is the
line-broadening of the quantum dot $\Gamma=\Gamma_L+\Gamma_R=\pi (t_L^2+t_R^2)N(0)$ with
the density of states of the reservoirs $N(0)$.

Even in the non-interacting limit, the problem of combining the
density matrix and the time-evolution operator remains. To resolve
the issue, we extend the chemical potential into complex numbers such that the
new chemical potential does not alter the quantum statistics while the
time-evolution rate along the imaginary-time can be shifted.
We introduce an effective non-interacting Hamiltonian $\hat{K}_0$,
\begin{equation}
\hat{K}_0=\hat{H}_0+(i\varphi_m-\Phi)\hat{Y}_0,
\label{eq:k0}
\end{equation}
with the Matsubara voltage,
\begin{equation}
i\varphi_m=i\frac{4\pi m}{\beta}\mbox{ for any integer }m.
\end{equation}
With the addition of the Matsubara voltage, the density matrix given by
$\hat{K}_0$ becomes $e^{-\beta K_0}=e^{-\beta[H_0+(i\varphi_m-\Phi) Y_0]}$. Since
$[H_0,Y_0]=0$, the operator can be written as $e^{-\beta(H_0-\Phi
Y_0)}\cdot e^{-i\beta\varphi_m Y_0}$. With
respect to the Fock basis constructed from the non-interacting operators
$\psi^\dagger_{0,\alpha k\sigma}$, $H_0$ and $Y_0$ are diagonal and the
eigenvalues of $Y_0$ are half- or full-integers. Therefore, we have an
operator identity
$e^{-i\beta\varphi_m Y_0}=1$ and
\begin{equation}
e^{-\beta K_0}=e^{-\beta(H_0-\Phi Y_0)}=\hat{\rho}_0,
\label{eq:key}
\end{equation}
independent of $i\varphi_m$. This effective Hamiltonian amounts to a
electron system with the statistics given by $\hat{\rho}_0$ and the
time-evolution is controlled at an independent phase velocity
shifted by the \textit{pumping} frequency of $i\varphi_m$. This
extra complex term is only applied to the imaginary-time action and
should not be interpreted as a physical damping term in real-time
formalism.

Given the non-interacting Hamiltonian, Eq.~(\ref{eq:k0}), we introduce
the interacting Hamiltonian $\hat{K}$, parametrized by $i\varphi_m-\Phi$,
with the many-body interaction $\hat{V}$ as
\begin{equation}
\hat{K}=\hat{K}_0+\hat{V}=\hat{H}_0+(i\varphi_m-\Phi)\hat{Y}_0+\hat{V},
\label{eq:k}
\end{equation}
and treat this as an effective Hamiltonian to the equilibrium
imaginary-time theory. In the following section, we will show that, upon
the analytic continuation $i\varphi_m\to\Phi$ the thermal Green's functions
map to nonequilibrium real-time Green's functions.

The equivalence of the imaginary-time theory and the real-time Keldysh
formalism can be best shown in explicit perturbative expansions. The
real-time retarded Green's function for the quantum dot is defined as
\begin{eqnarray}
G^{ret}(t)&=&-i\theta(t)\langle \{ d_H(t),d_H^\dagger(0)\}\rangle\\
&=&\theta(t)[G^>(t)-G^<(t)],
\end{eqnarray}
with the lesser and greater Green's functions defined as
\begin{eqnarray}
G^>(t)&=&-i\langle d_H(t)d^\dagger_H(0)\rangle\\
G^<(t)&=& i\langle d^\dagger_H(0)d_H(t)\rangle.
\end{eqnarray}
The subscript $H$ refers to the evolution in the Heisenberg picture
with the full Hamiltonian, $d^\dagger(t)=e^{itH}d^\dagger e^{-itH}$.
$\theta(t)$ is the step-function.

\begin{figure}[bt]
\rotatebox{0}{\resizebox{3.3in}{!}{\includegraphics{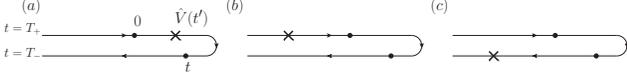}}}
\caption{(a) Keldysh contour for greater Green's function with the first order
scattering (marked by X) happening between time
$t=0$ and $t$ [corresponding to the case (i) considered in the text] (b)
Time ordering of case (ii). (c) Case (iii).
}
\label{fig:contour}\end{figure}

Usually an interaction picture is defined with a time-dependent
many-body interaction $\hat{V}$ turned on
adiabatically from the infinite past $(t=T\mbox{ as
}T\to -\infty$, see Fig.~\ref{fig:contour}).
The Green's functions can be defined in the Keldysh contour as
\begin{equation}
G^>(t)=-i Z_{0,neq}^{-1}{\rm Tr}\left[\hat{\rho}_0{\cal T}_K
e^{-i\int_K V_I(t')dt'}d(t)d^\dagger(0)\right].
\label{eq:int}
\end{equation}
The real-time evolution is given in the interaction picture as
$d^\dagger(t)=e^{itH_0}d^\dagger e^{-itH_0}$ and
$\hat{V}_I(t)e^{itH_0}\hat{V} e^{-itH_0}$. Time variables are defined on
the Keldysh contour, denoted as $K$. The nonequilibrium partition
function is defined as $Z_{0,neq}={\rm Tr}\hat{\rho}_0$.

Here, to avoid explicit time-dependence to the interaction $\hat{V}$,
we use Gell-Mann and Goldberger's~\cite{gellmann} construction of
steady-state when the limit $T\to -\infty$ is taken. Their
formalism uses full-strength interaction instead of adiabatic
interaction with the turn-on time
$T$ integrated from the remote past to the present. For example, a
scattering state operator defined as
\begin{equation}
\psi^\dagger_k=\eta\int_{-\infty}^0 dT\, e^{\eta T}e^{-i{\cal
L}T}e^{i{\cal L}_{0c}
T}c^\dagger_k,
\end{equation}
satisfies the Lippman-Schwinger equation
\begin{equation}
\psi^\dagger_k=c^\dagger_k+\frac{1}{\epsilon_k-{\cal
L}+i\eta}[\hat{V}+\hat{H}_{0t},c^\dagger_k],
\end{equation}
with the tunneling part of the non-interacting Hamiltonian
$\hat{H}_{0t}$. Here the infinitesimal $\eta$ determines the direction
of the time propagation.
The Liouville operators are defined as
${\cal L}\hat{A}=[\hat{H},\hat{A}]$, ${\cal
L}_{0c}\hat{A}=[\hat{H}_{0c},\hat{A}]$,
and $e^{-i{\cal L}T}\hat{A}
=e^{-i\hat{H}T}\hat{A}e^{i\hat{H}T}$, etc.
Here the adiabatic limit $T\to -\infty$ is replaced by an integral
via $\eta\int_{-\infty}^0 dT\,e^{\eta T}$ where the start
time $T$ of interaction is averaged over the whole range 
$(-\infty,0]$. This averaging effectively cancels out the transient
phase of each scattered wave which has propagated with the interaction
turned on at time $T$.

The greater Green's function $G^>(t)$ can be represented as
Fig.~\ref{fig:contour} with a $d$-electron created at time $0$ on the
upper contour and an electron destroyed at time $t$ on the lower
contour. Then, the first order perturbation
correction from Eq.~(\ref{eq:int}) can be decomposed into three
different time-ordered terms
according to the time of interaction $t'$ as (i) $0_+<t'<t_-$ (ii)
$T_+<t'<0_+$ (iii) $t_-<t'<T_-$, as shown in
Figs.~\ref{fig:contour}(a)-(c), respectively. The contribution for (i) can be expressed
as (with $Z_{0,neq}^{-1}$ omitted for brevity)
\begin{eqnarray}
&&-i\int_0^t \left\langle
e^{-i(T-t)H_0}d e^{-i(t-t')H_0}\hat{V}e^{-it'H_0}
d^\dagger e^{iTH_0}
\right\rangle_0 dt' \nonumber \\
& = &-i\int_0^t \left\langle
e^{itH_0}d e^{-i(t-t')H_0}\hat{V}e^{-it'H_0}
d^\dagger\right\rangle_0 dt' \nonumber \\
& = &-i\int_0^t \sum_{nmk}
\hat{\rho}_{0,n} e^{it(E_n-E_m)}d_{nm} e^{it'(E_m-E_k)}\hat{V}_{mk}
d^\dagger_{kn}dt'\nonumber \\
& = &-\sum_{nmk}
\hat{\rho}_{0,n} \left[e^{it(E_n-E_m)}-e^{it(E_n-E_k)}\right]
\frac{d_{nm}\hat{V}_{mk}d^\dagger_{kn}}{E_m-E_k},
\label{eq:i}
\end{eqnarray}
where the states denoted by $n,m,k$ are Fock states constructed from the
non-interacting scattering states. $\langle\cdots\rangle_0$ is defined
as ${\rm Tr}[\hat{\rho}_0\cdots]$. Note that zero of the energy
denominator $E_m-E_k$ does not lead to a singularity since
$e^{it(E_n-E_m)}-e^{it(E_n-E_k)}=0$.

For the perturbation occurring in the interval extending to the
infinity on the upper time contour [Fig.~\ref{fig:contour}(b)], 
the integral for (ii) becomes
\begin{eqnarray}
&&-i\eta\int_{-\infty}^0dT\,e^{\eta T}
\int_{T}^0 dt'\left\langle
e^{itH_0}d e^{-itH_0}
d^\dagger e^{it'H_0}\hat{V}e^{-it'H_0}
\right\rangle_0 \nonumber \\
& &=-\sum_{nmk}
\hat{\rho}_{0,n} e^{it\Delta E_{nm}}
\frac{d_{nm}d^\dagger_{mk}\hat{V}_{kn}}{\Delta E_{kn}-i\eta},
\label{eq:ii}
\end{eqnarray}
with $\Delta E_{nm}=E_n-E_m$.
Similarly for the case (iii) with the interaction on the lower 
contour, we have
\begin{eqnarray}
&&-i\eta\int_{-\infty}^0dT\,e^{\eta T}
\int_{t}^{T} dt'\left\langle
e^{it'H_0}\hat{V}e^{-i(t'-t)H_0}d e^{-itH_0}
d^\dagger
\right\rangle_0 \nonumber \\
& &=\sum_{nmk}
\hat{\rho}_{0,n} e^{it\Delta E_{nk}}
\frac{\hat{V}_{nm}d_{mk}d^\dagger_{kn}}{\Delta E_{nm}-i\eta}.
\label{eq:iii}
\end{eqnarray}
By denoting the time ordering on the Keldysh contour as $(ab)_K$ for an
event $a$ following $b$, the Eqs.~(\ref{eq:i}-\ref{eq:iii}) can be
represented by a cyclic permutation of $\{(d V d^\dagger)_K,(d
d^\dagger V)_K,(Vd d^\dagger)_K\}$, respectively. Rearranging the
indices in Eqs. (\ref{eq:ii}-\ref{eq:iii}), the contributions from (ii)
and (iii) combine to
\begin{equation}
\sum_{nmk}
\left[\hat{\rho}_{0,n} e^{it\Delta E_{nk}}
-\hat{\rho}_{0,m} e^{it\Delta E_{mk}} \right]
\frac{\hat{V}_{nm}d_{mk}d^\dagger_{kn}}{\Delta E_{nm}-i\eta}.
\label{eq:shell}
\end{equation}
Finally the first-order perturbation to the greater Green's function becomes
\begin{eqnarray}
& &
-\sum_{nmk}
\hat{\rho}_{0,n} \left[e^{it\Delta E_{nm}}-e^{it\Delta E_{nk}}\right]
\frac{d_{nm}\hat{V}_{mk}d^\dagger_{kn}}{\Delta E_{mk}} \nonumber \\
&& 
+\sum_{nmk}
\left[\hat{\rho}_{0,n} e^{it\Delta E_{nk}}
-\hat{\rho}_{0,m} e^{it\Delta E_{mk}} \right]
\frac{\hat{V}_{nm}d_{mk}d^\dagger_{kn}}{\Delta E_{nm}-i\eta}.
\end{eqnarray}

We perform the same perturbation theory to the thermal Green's function
defined with the imaginary-time under the Hamiltonian $\hat{K}$,
Eq.~(\ref{eq:k}), as
\begin{equation}
{\cal G}(t)=-\langle {\cal T}_\tau d_K(\tau)d_K^\dagger(0)\rangle,
\end{equation}
with the time-evolution given as
$d_K(\tau)=e^{\tau K}d e^{-\tau K}$. Using the interaction picture, the
Green's function is expanded in a perturbation series as
\begin{equation}
{\cal G}(t)=- Z_{0,neq}^{-1}{\rm Tr}\left[\hat{\rho}_0{\cal T}_\tau
e^{-\int_0^\beta V_I(\tau')d\tau'}d(\tau)d^\dagger(0)\right],
\end{equation}
where the time-evolution in the interaction picture is given as
$d(\tau)=e^{\tau K_0}d e^{-\tau K_0}$. For $\tau>0$, the first perturbation
contribution due to the scattering of $\hat{V}$ at $\tau'$ can be
grouped into two cases; (i) $0<\tau'<\tau$ and (ii) $\tau<\tau'<\beta$.

The contribution from (i) $0<\tau'<\tau$ is
\begin{eqnarray}
&&-\int_0^\tau {\rm Tr}\left[
e^{-(\beta-\tau)K_0}d e^{-(\tau-\tau')K_0}
\hat{V}e^{-\tau'K_0}d^\dagger
\right]d\tau' \nonumber \\
& = &\sum_{nmk}
\rho_{0,n}\left[
e^{\tau\Delta K_{nm}}-e^{\tau\Delta K_{nk}}\right]
\frac{d_{nm}\hat{V}_{mk}d^\dagger_{kn}}{\Delta K_{mk}},
\end{eqnarray}
with the eigenvalues $K_n$ of $\hat{K}_0$ and $\Delta K_{nm}=K_n-K_m$.
Here we have used the key relation Eq.~(\ref{eq:key}), $e^{-\beta
\hat{K}_0}=\hat{\rho}_0$. If the analytic continuation
$i\varphi_m\to\Phi$ is formally carried out followed by $\tau\to it$,
$K_n\to E_n$ and
the above integral for $0<\tau'<\tau$ becomes identical to the real-time Green's function
Eq.~(\ref{eq:i}) for $0_+<t'<t_-$.
Now, for the case of (ii) $\tau<\tau'<\beta$, the integral becomes
\begin{eqnarray}
&&-\int_\tau^\beta {\rm Tr}\left[
e^{-(\beta-\tau')K_0}\hat{V}e^{-(\tau'-\tau)K_0}d e^{-\tau K_0}
d^\dagger
\right]d\tau' \nonumber \\
& = &\sum_{nmk}
\left[\rho_{0,n}e^{\tau\Delta K_{nk}}-\rho_{0,m}
e^{\tau\Delta K_{mk}}\right]
\frac{\hat{V}_{nm}d_{mk}d^\dagger_{kn}}{\Delta K_{nm}},
\end{eqnarray}
which transforms to the real-time result, Eq.~(\ref{eq:iii}), with the
analytic continuation except for the adiabatic factor $i\eta$.

We discuss the subtlety of the adiabatic factor $i\eta$,
the presence of which is only relevant for the energy shell of
$E_n=E_m$ in Eq.~(\ref{eq:shell}),
\begin{equation}
i\pi\sum_{nmk}\left[
\hat{\rho}_{0,n}-\hat{\rho}_{0,m}\right]e^{it\Delta E_{nk}}
\hat{V}_{nm}d_{mk}d^\dagger_{kn}\delta(E_n-E_m).
\label{eq:eta}
\end{equation}
In equilibrium, $\hat{\rho}_0$ is only given by energy and the above
expression is zero since $\rho_{0,n}-\rho_{0,m}=0$. To extend the
argument to nonequilibrium,
we take an explicit example of one-quantum dot with a two-body
interaction such as the on-site Coulomb interaction
$\hat{V}=Un_{d\uparrow}n_{d\downarrow}$. In terms of the explicit
scattering state operators Eq.~(\ref{eq:psi}),
\begin{eqnarray}
d^\dagger_\sigma&=&\sum_{\alpha k}\frac{t_\alpha}{\sqrt\Omega}g^*(\epsilon_{\alpha
k})\psi^\dagger_{\alpha k\sigma} \\
\hat{V}&=&U\sum_{\{\alpha, k\}}
\left(\frac{t_{\alpha_1}g^*(\epsilon_1)}{\sqrt\Omega}\psi^\dagger_{\alpha1
k1\uparrow}\right)
\left(\frac{t_{\alpha_2}g(\epsilon_2)}{\sqrt\Omega}\psi_{\alpha2
k2\uparrow}\right)\nonumber \\
& & \times
\left(\frac{t_{\alpha_3}g^*(\epsilon_3)}{\sqrt\Omega}\psi^\dagger_{\alpha3
k3\downarrow}\right)
\left(\frac{t_{\alpha_4}g(\epsilon_4)}{\sqrt\Omega}\psi_{\alpha4
k4\downarrow}\right).
\label{eq:vpsi}
\end{eqnarray}

\begin{figure}[bt]
\rotatebox{0}{\resizebox{2.3in}{!}{\includegraphics{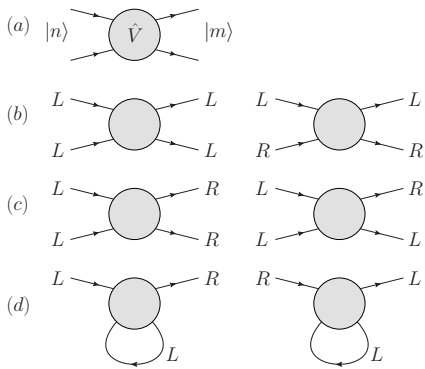}}}
\caption{(a) Interaction $\hat{V}$ mapping a state $|n\rangle$ to
$|m\rangle$. (b) Initial and final states without changes in the
$Y_0$-number of scattering states. 
(c) Initial and final states with different
$Y_0$-numbers. (d) Contraction with the $d$-creation and annihilation
operators. Each line represents the contraction of
scattering state basis $\langle \psi^\dagger_{\alpha k\sigma}
\psi_{\alpha k\sigma}\rangle_0$ and contributes the factor
$t_\alpha^2|g(\epsilon_{\alpha k})|^2$. Two diagrams are with
$|n\rangle$ and $|m\rangle$ states interchanged.
}
\label{fig:subtle}\end{figure}

\begin{figure}[bt]
\rotatebox{0}{\resizebox{3.1in}{!}{\includegraphics{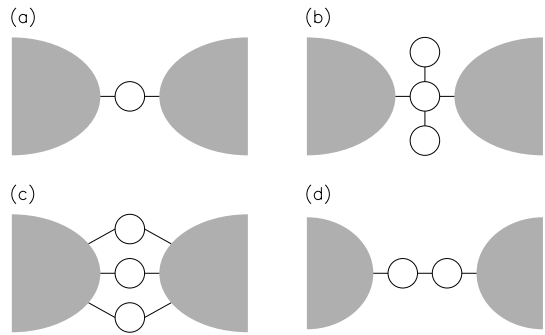}}}
\caption{Nonlinear Transport through (a) single QD, (b) side-coupled QDs
and (c) parallel QDs can be studied within the imaginary-time formalism.
(d) Green's functions in serially coupled QDs may not be correctly
continued to lesser Green's functions.
}
\label{fig:qd}\end{figure}

As depicted in the Fig.~\ref{fig:subtle}(b), if the incoming and
outgoing states $|n\rangle$ and $|m\rangle$ conserve the eigenvalues
for $\hat{Y}_0$ operator ($Y_0$-number), the factor
$(\hat{\rho}_{0,n}-\hat{\rho}_{0,m})$ is zero. Figure~\ref{fig:subtle}(c)
shows the terms in $\hat{V}$ [Eq.~(\ref{eq:vpsi})] which do not conserve the $Y_0$-number.
However, when expectation values are taken 
between the states $|n\rangle$ and $|m\rangle$, it can be shown that
there is a counter-term which leads to a cancellation. As demonstrated
in Fig.~\ref{fig:subtle}(d), the connected legs represent the contraction
$\langle \psi^\dagger_{\alpha k\sigma}
\psi_{\alpha k\sigma}\rangle_0$ and contribute the factor
$t_\alpha^2|g(\epsilon_{\alpha k})|^2$ from
Eqs.~(\ref{eq:psi}) and (\ref{eq:vpsi}).
If the source and drain reservoirs are given by the same continuum
density of states, there is always a contribution with the same magnitude
from when the states
$|n\rangle$ and $|m\rangle$ are interchanged, therefore leading to the
cancellation of the
factor $(\hat{\rho}_{0,n}-\hat{\rho}_{0,m})$. We emphasize that the
above argument does not require assumptions for the source-drain
symmetry $t_L=t_R$ or the particle-hole symmetry. It also holds for
different types of on-site interaction.

To summarize, the time-ordering of the real-time greater Green's function
can be matched to the imaginary-time-ordering [denoted by $(\cdots)_I$] as
$(dVd^\dagger)_K\leftrightarrow(dVd^\dagger)_I$ and
$(Vdd^\dagger)_K+(dd^\dagger V)_K\leftrightarrow(Vdd^\dagger)_I$.
The higher order contributions can be checked in the similar manner to
the first order. For example, in the second order, the
time-orderings in the real-time theory can be matched as
$(dVVd^\dagger)_K\leftrightarrow(dVVd^\dagger)_I$,
$(VdVd^\dagger)_K+(dVd^\dagger V)_K\leftrightarrow(VdVd^\dagger)_I$, and
$(VVdd^\dagger)_K+(Vdd^\dagger V)_K+(dd^\dagger VV)_K\leftrightarrow(VVdd^\dagger)_I$.
Such topologically equivalent graphs between the imaginary-time and real-time expansions
at each perturbation order are
expected since the two theories are known to be equivalent in
equilibrium.

The lesser Green's function can be shown to be equivalent to the imaginary-time
Green's function ${\cal G}(\tau)$ for $\tau<0$.
As shown previously~\cite{prl07}, if Fourier transformation is
performed on the real- and imaginary-time Green's functions, the retarded
Green's function $G^{ret}(\omega)$ can be obtained by analytically
continuing the thermal Green's function ${\cal G}(i\omega_n)$ via
$i\omega_n\leftrightarrow \omega+i\eta$.

So far, we discussed the analytic continuation in single-QD structures
[Fig.~\ref{fig:qd}(a)]. However, the formulation can be
straightforwardly extended to much wider range of quantum dot structures
such as the side-coupled QD and parallel-coupled QD systems as shown in
Fig.~\ref{fig:qd}(b-c). For serially coupled QD systems
[Fig.~\ref{fig:qd}(d)], the current imaginary-time formulation does not
have an analytical continuation to real-time Green's functions and hence
the formulation in this work cannot be applied.

\section{Computational Algorithm}
\label{sec:algorithm}

We implement the imaginary-time formalism for a numerical nonequilibrium
technique using
quantum Monte Carlo method. We use the Hirsch-Fye (HF)
algorithm~\cite{hirsch} which
uses only the interacting orbitals (QD site) as the dynamic variable
after integrating out the non-interacting reservoir electronic states. The
QD Green's function is stochastically updated along the discretized
imaginary-time lattice using the local update method by Blankenbecler
et al~\cite{bss}. Calculation of nonequilibrium interacting Green's 
function does not
require any main modifications of the standard (equilibrium) QMC code.

\subsection{Non-interacting Green's function}
Independent sets of simulations are performed with different
$i\varphi_m$'s in Eq.~(\ref{eq:k})
treated as fixed parameters to QMC. In the
following calculations, $0\leq m\leq 5$ have been used. Given
$i\varphi_m$, the non-interacting QD Green's function can be easily derived as
\begin{eqnarray}
{\cal G}^0(i\omega_n) & = &
\sum_{\alpha k}\langle d_\sigma|\psi_{\alpha k\sigma}\rangle
\frac{1}{i\omega_n-K_{0,\alpha k}}
\langle\psi_{\alpha k\sigma}|d_\sigma\rangle \\
& = & \sum_{\alpha k}\frac{t_\alpha^2}{\Omega}
\frac{|g(\epsilon_{\alpha k})|^2}{i\omega_n-\epsilon_{\alpha
k}-\alpha\frac{i\varphi_m-\Phi}{2}}
\label{eq:gw0}\\
&=&\sum_\alpha\frac{\Gamma_\alpha/\Gamma}{
z_{nm}+i\Gamma\cdot {\rm Sign}({\rm Im}\,z_{nm})},
\end{eqnarray}
with $z_{nm}=i\omega_n-\alpha\frac{i\varphi_m-\Phi}{2}$.
If the analytic continuation of $i\varphi_m\to\Phi$ followed by 
$i\omega_n\to\omega+i\eta$ is performed, the thermal Green's function in the Matsubara
frequency transforms to the real-time retarded Green's function,
\begin{equation}
\sum_\alpha\frac{\Gamma_\alpha/\Gamma}{
\omega+i\eta+i\Gamma}=\frac{1}{\omega+i\eta+i\Gamma}.
\end{equation}
Fourier transformation to the imaginary-time variable ($\tau>0$)
gives
\begin{eqnarray}
{\cal G}^0(\tau) & = & \frac{1}{\beta}\sum_n {\cal
G}^0(i\omega_n)e^{-i\omega_n\tau}\\
& = &-\sum_{\alpha k}\frac{t_\alpha^2}{\Omega}|g^0(\epsilon_{\alpha
k})|^2e^{-\tau[\epsilon_{\alpha k}+\alpha(i\varphi_m-\Phi)/2]}
[1-f_\alpha(\epsilon)],
\label{eq:g0}
\end{eqnarray}
with the Fermi-Dirac function in the $\alpha$-reservoir,
$f_\alpha(\epsilon)=[1+e^{\beta(\epsilon-\alpha\Phi/2)}]^{-1}$.
This expression is later used as the input to the QMC calculation.
If we perform the analytic continuation $i\varphi_m\to\Phi$ followed by
$\tau\to it$, the Green's function in the imaginary-time transforms to
\begin{equation}
-\sum_{\alpha k}\frac{t_\alpha^2}{\Omega}|g^0(\epsilon_{\alpha 
k})|^2e^{-it\epsilon_{\alpha k}}[1-f_\alpha(\epsilon_{\alpha k})],
\end{equation}
which is nothing but the non-interacting greater Green's function of 
QD orbital, $G^>(t)$, in the
steady-state nonequilibrium. In this
sense, the thermal Green's function before analytic continuations contains
full information of the retarded, greater and lesser Green's functions in the
real-time formulation.

In this work, we analytically continue the retarded (self-energy)
functions instead of the lesser/greater Green's functions as functions
of frequency since
the retarded functions have simpler analytic structure of 
rational functions, as opposed to the exponential functions for the
lesser/greater Green's functions as functions of time.

\subsection{Quantum Monte-Carlo method and Self-energy}

The QMC procedure follows the standard Hirsch-Fye algorithm
with the initial Green's function Eq.~(\ref{eq:g0}) at fixed $i\varphi_m$.
QMC method stochastically samples the fermionic phase space via
auxiliary fields in the action introduced by the
Hubbard-Stratonovich transformation~\cite{hirsch}. The 
auxiliary fields are updated according to the effective Boltzmann
factor~\cite{bss}. The only modification to the HF algorithm
is that the Monte Carlo (MC) Green's function
$G(\tau,\tau')$ at
an auxiliary field configuration is complex in contrast to the
equilibrium calculations. Since $G(\tau,\tau')$ is complex, the
ratio of Boltzmann factors for the new and old auxiliary field
configurations is also complex in general. Therefore for any observable
$\langle \hat{A}\rangle$ we compute the ensemble average as
\begin{equation}
\langle \hat{A}\rangle = \frac{\sum_n f(n)A(n)}{\sum_n f(n)}
=\frac{\sum_n e^{i\theta_n}A(n)|f(n)|}{\sum_n e^{i\theta_n}|f(n)|}
=\frac{\langle\!\langle e^{i\theta}A\rangle\!\rangle}{
\langle\!\langle e^{i\theta}\rangle\!\rangle},
\end{equation}
with the effective Boltzmann factor $f(n)$ for a auxiliary field
configuration $n$ and its phase factor $e^{i\theta_n}=f(n)/|f(n)|$. The
ensemble average $\langle\!\langle \cdots\rangle\!\rangle$ is taken over
the Markov chain of the Monte Carlo configurations chosen by the probability $|f(n)|$.
In the calculations shown later, the average phase factor
$|\langle\!\langle e^{i\theta}\rangle\!\rangle|$ remained 
close to one (typically $0.9-1.0$) and the statistics has been quite
robust. We note that at $\Phi=0$ and $i\varphi_m=0$, the QMC calculation
is completely identical to the equilibrium QMC method.

\begin{figure}[bt]
\rotatebox{0}{\resizebox{3.4in}{!}{\includegraphics{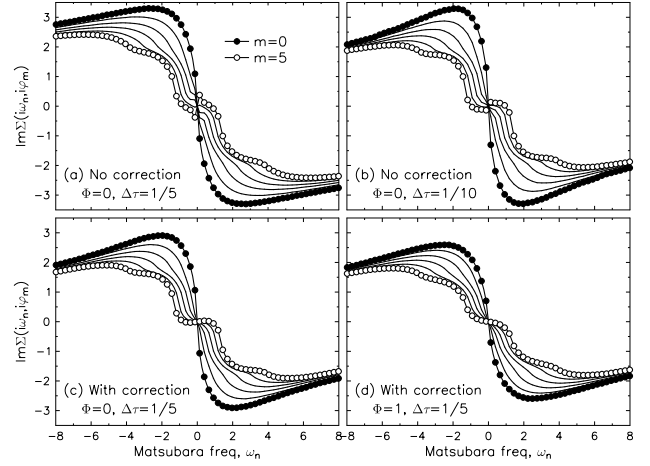}}}
\caption{Imaginary-time electron self-energy at $U=10$ and $\beta=24$
for Matsubara voltages $\varphi_m$ with $m=0\mbox{ (filled
circle)},\cdots,5\mbox{ (open circle)}$.
(a) Conventional Hirsch-Fye algorithm without the discretization
correction shows spurious structures at small $\omega_n$ and finite
$\varphi_m$. (b) With smaller discretization $\Delta\tau=1/10$, the
spurious structures become weaker. (c) The discretization
correction produced smooth self-energy at small $\omega_n$. (d) At
higher bias $\Phi=1$, the curvature at high $\varphi_m$ becomes weaker,
suggesting reduced correlation effects. The unit of energy is given by
the non-interacting level width of QD, $\Gamma=1$.
}
\label{fig:sigma}\end{figure}

The QMC Green's function defined on a discrete imaginary-time mesh
$\tau_i=i\Delta\tau (\Delta\tau=\beta/N,\, i=0,\cdots,N-1)$ is updated by the QMC Dyson's
equation~\cite{bss,hirsch}
\begin{equation}
{\bf G}={\bf G}_0+({\bf G}_0-I)(e^{\bf V}-I){\bf G},
\end{equation}
for the Green's function matrix defined as ${\bf G}_{ij}=G(\tau_i,\tau_j)$.
Here ${\bf V}$ represents the auxiliary-field coupling to electrons after the
Hubbard-Stratonovich transformation~\cite{hirsch,bss} of many-body
interaction.
The diagonal component of the matrix ${\bf G}$ is chosen as the greater
Green's function by~\cite{negele}
\begin{equation}
G_{ii}=G^>(\tau_i,\tau_i)=G(\tau_i+0^+,\tau_i).
\end{equation}
This choice of zero-time Green's function introduces discretization error
when the time variables are integrated in the above Dyson's equation or
in other observables. In nonequilibrium
calculations, the Green's function, Eq.~(\ref{eq:g0}), has an additional oscillation due to the 
$i\varphi_m$ dependence, and the
systematic error of discretization becomes more significant. Spurious structures in the
low frequency self-energy observed in Ref~\cite{prl07} are attributed
to the discretization error~\cite{dirks}, which can be confirmed in
comparison with the continuous-time QMC~\cite{rubtsov,gull}. In
Figs.~\ref{fig:sigma}(a-b), the discretization error is compared
for $\Delta\tau=1/5$ and $1/10$. The energy unit is the non-interacting
broadening $\Gamma$. Although the self-energy at high frequency
$\omega_n$ is well convergent, the low frequency
$\Sigma(i\omega_n,i\varphi_m)$ shows discontinuous jumps at
$\omega_n\approx 0$ as $\varphi_m$ increases. This kink becomes smoother
as $\Delta\tau$ becomes small. Without the correction, the analytic
continuation misinterprets the kink in the self-energy due to
incoherent spectra and exaggerated the destruction of Kondo resonance at
finite bias~\cite{prl07}.

By adopting
a similar trick for Fourier transformation considered in the
continuous-time QMC~\cite{rubtsov,gull}, 
the Green's functions in the discrete-time QMC has
been measured as follows. When involved in a time-integral, we use the non-interacting 
Green's function matrix $\tilde{\bf G}_0$ with the diagonal
elements augmented by $\tilde{\bf G}_{0,ii}=\frac12\left[G_0(\tau_i+0^+,\tau_i)
+G_0(\tau_i-0^+,\tau_i)\right]=G_0(\tau_i+0^+,\tau_i)-\frac12$, \textit{i.e.}
$\tilde{\bf G}_0={\bf G}_0-\frac12 I$. Then the Dyson's equation can be
rewritten as
\begin{equation}
{\bf G}={\bf G}_0+\tilde{\bf G}_0{\bf S}\tilde{\bf G}_0
+\frac12\left({\bf S}\tilde{\bf G}_0-\tilde{\bf G}_0{\bf S}\right),
\label{dyson1}
\end{equation}
with the $S$-matrix defined as
\begin{equation}
{\bf S}=(e^{\bf V}-I){\bf G}{\bf G}^{-1}_0.
\end{equation}
Through Monte Carlo updates we measure the Fourier transformed
$S$-matrix as
\begin{equation}
S(i\omega_n)=\left\langle\!\!\!\left\langle
\frac{1}{\beta}\sum_{ij}e^{i\omega_n(\tau_i-\tau_j)}(e^{V(i)}-1)[{\bf
G}{\bf G}^{-1}_0]_{ij}
\right\rangle\!\!\!\right\rangle.
\end{equation}
Here the matrix ${\bf G}$ is calculated at each update of the auxiliary
fields and ${\bf G}_0^{-1}$ is calculated and stored at the beginning of
computation.
The last term in Eq.~(\ref{dyson1}) vanishes due to the time
translational symmetry and 
\begin{equation}
{\cal G}(i\omega_n)={\cal G}_0(i\omega_n)+{\cal G}_0(i\omega_n)
{\cal S}(i\omega_n){\cal G}_0(i\omega_n).
\label{dyson2}
\end{equation}
As shown in Fig.~\ref{fig:sigma}(c), the unphysical structure at
$\omega_n\approx 0$ disappeared even at $\Delta\tau=1/5$ after the
discretization errors have been corrected. Figure~\ref{fig:sigma}(d) at a
finite bias $\Phi$ shows less curvature in the self-energy, which
suggests that the correlation effects become weaker as bias increases.
Green's functions evaluated this way showed excellent agreement with the
continuous-time QMC~\cite{dirks} at low fermion frequencies $i\omega_n$ at large
Matsubara frequencies $i\varphi_m$ with computationally accessible
$\Delta\tau$. In the calculations presented below used
$\Delta\tau=1/5$ unless mentioned otherwise.

The self-energy of the QD Green's function $\Sigma(i\omega_n,i\varphi_m)$
is then computed in the same manner
as in equilibrium theory, via the Dyson's equation 
\begin{equation}
\Sigma(i\omega_n,i\varphi_m)=\left[{\cal G}^0(i\omega_n)\right]^{-1}
-\left[{\cal G}(i\omega_n)\right]^{-1}.
\end{equation}
This self-energies $\Sigma(i\omega_n,i\varphi_m)$ at different
$i\varphi_m$ values are computed in separate sets of QMC runs at each
$i\varphi_m$. The numerical data for $\Sigma(i\omega_n,i\varphi_m)$ is
analytically continued to the retarded self-energy
$\Sigma^{ret}(\omega)$ with the real-frequency $\omega$. The numerical
procedure will be fully discussed in the next section.

We comment on why we choose to analytically continue the self-energy
instead of the Green's function directly. As will be clear in subsequent
discussions, the analytic form of the perturbative energy self-energy is
more readily written down, which makes the numerical
procedure more transparent. From the numerical standpoint of the
discrete-time QMC, the
discretization makes the high Matsubara frequency
data less reliable for $\Sigma(i\omega_n,i\varphi_m)$ and
${\cal G}(i\omega_n,i\varphi_m)$. However, the systematic errors in
an analytically continued $\Sigma(\omega)$ at large $\omega$ are less
problematic since the frequency term $\omega$ in the Dyson's equation
$G^{ret}(\omega)=[\omega-\epsilon_d+i\Gamma-\Sigma(\omega)]^{-1}$
dominates $\Sigma(\omega)$.

\section{Numerical Analytic continuation}
\label{sec:analytic}

\subsection{The spectral ansatz}

\begin{figure}[bt]
\rotatebox{0}{\resizebox{1.7in}{!}{\includegraphics{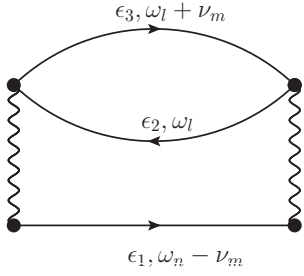}}}
\caption{Self-energy diagram of second order perturbation in the Coulomb
parameter $U$ of the Anderson model.
}
\label{fig:diagram}\end{figure}

It seems a formidable task to perform an analytic continuation on
numerical data in $\Sigma(i\omega_n,i\varphi_m)$. To guide the analytic
continuation to a correct form we start with the self-energy in the
second order of Coulomb interaction in the Anderson model, with the
diagram depicted in
Fig.~\ref{fig:diagram}. The retarded self-energy can be easily
calculated from $\Sigma^\gtrless(t)=U^2[G_0^\gtrless(t)]^2G_0^\lessgtr(-t)$
and $\Sigma^{ret}(t)=\theta(t)[\Sigma^>(t)-\Sigma^<(t)]$ with
the step-function $\theta(t)$,
\begin{eqnarray}
\Sigma^{ret}(\omega) & = &U^2\sum_{\alpha_1,\alpha_2,\alpha_3}\left[
\prod_{i=1}^3 
\frac{\Gamma_i}{\Gamma} \int d\epsilon_i
\rho_0(\epsilon_i)\right] \nonumber \\
& &\times\frac{
f_1(1-f_2)f_3+(1-f_1)f_2(1-f_3)
}{\omega+i\eta-\epsilon_1+\epsilon_2-\epsilon_3},
\end{eqnarray}
with the shorthand notation $f_i=f_{\alpha_i}(\epsilon_i)$.
$\rho_0(\epsilon)$ is the non-interacting QD spectral function. If we do the
same diagram in the imaginary-time formalism with the Hamiltonian
$\hat{K}$, the self-energy is
\begin{eqnarray}
\Sigma(i\omega_n,i\varphi_m) & = &U^2\sum_{\alpha_1,\alpha_2,\alpha_3}\left[
\prod_{i=1}^3 
\frac{\Gamma_i}{\Gamma} \int d\epsilon_i
\rho_0(\epsilon_i)\right] \nonumber \\
& &\times\frac{
f_1(1-f_2)f_3+(1-f_1)f_2(1-f_3)
}{i\omega_n-\tilde\epsilon_1+\tilde\epsilon_2-\tilde\epsilon_3},
\end{eqnarray}
with $\tilde\epsilon_i=\epsilon_i+\alpha_i(i\varphi_m-\Phi)/2$. Here we
have used the relation for the Fermi-Dirac function,
\begin{equation}
f\left(\epsilon+\alpha\frac{i\varphi_m-\Phi}{2}\right)
=
f\left(\epsilon-\alpha\frac{\Phi}{2}\right)
=f_\alpha(\epsilon),
\end{equation}
which is equivalent to Eq.~(\ref{eq:key}).
By combining the reservoirs indices
\begin{equation}
\gamma=\alpha_1-\alpha_2+\alpha_3
\end{equation}
and the energy of an electron dressed by an electron-hole pair as
$\epsilon=\epsilon_1-\epsilon_2+\epsilon_3$, we can rewrite the above
expression as
\begin{equation}
\Sigma(i\omega_n,i\varphi_m) = 
\sum_{\gamma=\pm 1,\pm 3}\int d\epsilon\frac{\sigma_\gamma(\epsilon)}{
i\omega_n-\frac{\gamma}{2}(i\varphi_m-\Phi)-\epsilon},
\label{eq:ansatz}
\end{equation}
with the spectral function $\sigma_\gamma(\epsilon)$ defined as
\begin{eqnarray}
\sigma_\gamma(\epsilon)
&=&\pi{U^2}\sum_{\alpha_1,\alpha_2,\alpha_3}^{\alpha_1-\alpha_2+\alpha_3=\gamma}\left[
\prod_{i=1}^3
\frac{\Gamma_i}{\Gamma} \int d\epsilon_i
\rho_0(\epsilon_i)\right] \nonumber \\
& &\times[
f_1(1-f_2)f_3+(1-f_1)f_2(1-f_3)] \nonumber \\
& & \times
\delta(\omega-\epsilon_1+\epsilon_2-\epsilon_3).
\end{eqnarray}
In the second order of interaction the self-energy spectral function
$\sigma_\gamma(\epsilon)$ is independent of $i\varphi_m-\Phi$. However
in the higher order of perturbation, it is no longer the case and we
need to incorporate the $i\varphi_m-\Phi$ dependence in the spectral
function as
\begin{equation}
\Sigma(i\omega_n,i\varphi_m) = 
\sum_{\gamma={odd\,Z}}\int
d\epsilon\frac{\sigma_\gamma(\epsilon)Q_\gamma(\epsilon,i\varphi_m-\Phi)}{
i\omega_n-\frac{\gamma}{2}(i\varphi_m-\Phi)-\epsilon},
\label{eq:spec1}
\end{equation}
with any odd integer $\gamma$. $Q_\gamma(\epsilon,i\varphi_m-\Phi)$ is
the correction due to higher order diagrams. See the appendix for detailed
discussions for this generalization of the spectral representation and
its analytic properties.
We approximate the function $Q_\gamma$ by a Pad\'e approximant
\begin{equation}
Q_\gamma(\epsilon,z)
=\frac{1+C_\gamma^{(1)}(\epsilon)z+C_\gamma^{(2)}(\epsilon)z^2+\cdots}{
1+D_\gamma^{(1)}(\epsilon)z+D_\gamma^{(2)}(\epsilon)z^2+\cdots}.
\label{eq:spec2}
\end{equation}
Therefore we seek the best spectral representation
of the QMC-computed self-energy by treating
$\{\sigma_\gamma(\epsilon),C_\gamma^{(n)}(\epsilon),D_\gamma^{(n)}(\epsilon)\}$
as fitting parameters. In the following calculations, we limit the
$\gamma$-branches to $\gamma=\pm1, \pm3, \pm5,\pm7$ and the Pad\'e
approximants to the first order $n=1$, which already required fitting 24
functions simultaneously. We will discuss in the next section the effects of the Pad\'e
approximants.
We emphasize that, although the Pad\'e coefficient functions
$\{C_\gamma^{(n)}(\epsilon),D_\gamma^{(n)}(\epsilon)\}$ are adjusted in
the numerical fit, they do not contribute to the (real-frequency)
self-energy after the analytic continuation $i\varphi_m\to\Phi$,
\begin{equation}
{\rm Im}\Sigma^{ret}(\omega)=-\pi\sum_\gamma \sigma_\gamma(\omega).
\end{equation}
Real part of $\Sigma^{ret}(\omega)$ is obtained from the Kramers-Kronig
relation.

The electron self-energy satisfies the general relation
\begin{equation}
\Sigma(i\omega_n,i\varphi_m)=[\Sigma(-i\omega_n,-i\varphi_m)]^*,
\end{equation}
as can be seen in the non-interacting Green's function,
Eq.~(\ref{eq:gw0}).
For a particle-hole symmetric system, the non-interacting Green's function
${\cal G}_0(i\omega_n,i\varphi_m)$ in Eq.~(\ref{eq:gw0}) is invariant
with $i\varphi_m-\Phi\leftrightarrow -i\varphi_m+\Phi$ and
we derive symmetry relations
\begin{eqnarray}
\sigma_\gamma(\epsilon)&=&\sigma_{-\gamma}(-\epsilon) \nonumber \\
Q_\gamma(\epsilon,i\varphi_m-\Phi)&=&
Q_{-\gamma}(-\epsilon,i\varphi_m-\Phi).
\label{eq:sym}
\end{eqnarray}
In the previous work~\cite{prl07}, the relation
$\sigma_\gamma(\epsilon)=\sigma_{-\gamma}(-\epsilon)$ has been
incorrectly applied as
$\sigma_\gamma(\epsilon)=\sigma_{-\gamma}(\epsilon)$ and this led to
overly rapid reduction of the Kondo resonance at finite bias.
As pointed out~\cite{prb10} later, correct
constraint produced a good agreement with other method~\cite{anders}
in the moderately interacting limit, $U=5$ [see Fig.~2(d) in
Ref.~\cite{prb10}]. In this work, we
do not impose any symmetry relations in the fit, and the resulting
spectral functions recovered the above relations numerically.

\subsection{Fitting procedures}
With the above spectral ansatz, we perform the least-square fit to the
numerical self-energy generated by the QMC calculations with 
$\chi^2$ defined as
\begin{equation}
\chi^2={\cal N}^{-1}\sum_{n=-N}^{N-1}\sum_{m=0}^M\sigma_{nm}^{-2}\left|
\frac{\Delta\Sigma(i\omega_n,i\varphi_m)}{
\Sigma_{QMC}(i\omega_n,i\varphi_m)}
\right|^2.
\end{equation}
The deviation between the self-energy fit
$\Sigma_{fit}(i\omega_n,i\varphi_m)$ in the above ansatz and the QMC 
generated data $\Sigma_{QMC}(i\omega_n,i\varphi_m)$ is
$\Delta\Sigma(i\omega_n,i\varphi_m)=\Sigma_{fit}(i\omega_n,i\varphi_m)
-\Sigma_{QMC}(i\omega_n,i\varphi_m)$. We fit more accurately the low
frequency self-energy data with an effective cutoff function
\begin{equation}
\sigma_{nm}^{-2}=\frac{\Gamma^2}{\omega_n^2+\Gamma^2}
\frac{\Gamma^2}{\varphi_m^2+\Gamma^2}.
\end{equation}
The normalization factor ${\cal N}$ is defined as
\begin{equation}
{\cal N}=\sum_{n=-N}^{N-1}\sum_{m=0}^M\sigma_{nm}^{-2}.
\end{equation}
In the following calculations we used $M=5$ and $\omega_N=8.0\Gamma$.

In QMC applications, the analytic continuation has been one of the most
controversial topics. Since the transformation of imaginary-time data to
real-time information is an ill-defined procedure, noise in the
imaginary-time data can lead to severe uncertainties in spectral
functions. In the past, several algorithms have been proposed and the
maximum entropy method based on the Bayesian inference~\cite{mem} and
the method of stochastic image generation~\cite{mishchenko} have been widely used.
In this work, we have not incorporated such methods where the
focus so far has been limited to finding right spectral representations.
Simultaneously finding a fit to many functions (24 as noted above) has
already been quite an extensive task computationally. Refining the analytic continuation
method remains an important area of future improvement for the imaginary-time
theory of nonequilibrium.

We discretize the frequency on logarithmic mesh systems with
201 frequency points centered at $\omega=0$ over the energy range
$[-30,30]$.
Minimization of $\chi^2$ is achieved iteratively by using the Newton's
steepest gradient method~\cite{payne}. Once the fit reaches a certain
threshold of accuracy ($\sqrt{\chi^2}<0.06$), we
regularized the spectral functions through third-order polynomial
smoothing to reduced unwanted noise. The smoothing has had mostly 
insignificant effects and often has been
unnecessary. The fractional error $\sqrt{\chi^2}$ in the fit resulted in
the range of $1-6$ \%. Generally, the Pad\'e approximant term produced better
fits. Due to the dense frequency points near $\omega=0$, the update of
spectral functions at small frequencies tends to be very slow. Therefore, for
faster convergence, we used adjustable mesh systems which evolved from a
coarse to a dense frequency mesh as the iterations progressed.

\subsection{Calculation of conductance}

Once the retarded Green's function is obtained, the current in a
single-quantum model can be computed from the Meir-Wingreen's
formula~\cite{wingreen},
\begin{equation}
I=\frac{e}{\hbar}\Gamma\int d\epsilon A(\epsilon,\Phi)[f_L(\epsilon)-f_R(\epsilon)],
\label{eq:meir}
\end{equation}
with the QD spectral function at the bias $\Phi$,
\begin{equation}
A(\epsilon,\Phi)=
-\frac{1}{\pi}{\rm Im}\,\frac{1}{\omega-\epsilon_d-\Sigma^{ret}(\omega)}.
\end{equation}
The differential conductance $G$ is obtained from numerical differentiation
of the current by
\begin{equation}
G=e\frac{dI}{d\Phi}.
\end{equation}
The differentiation has to be taken on discrete values of $\Phi_i$
$(i=0,1,\cdots)$. For
$\Phi_i=0$, the linear conductance is obtained from
\begin{equation}
G(\Phi=0)/G_0=\pi\Gamma \int d\epsilon A(\epsilon,0)\left(-\frac{\partial
f(\epsilon)}{\partial \epsilon}\right),
\end{equation}
with the conductance quantum $G_0$ given by
\begin{equation}
G_0=\frac{2e^2}{h}.
\end{equation}
For the first non-zero bias ($i=1$), we evaluate the derivatives from a
third-order polynomial at $\Phi=\Phi_1$ determined from the values of current 
$\{-I_{2},-I_{1},0,I_1,I_2\}$ at bias
$\{-\Phi_{2},-\Phi_{1},0,\Phi_1,\Phi_2\}$ as,
\begin{equation}
G_1=\frac13\frac{I_1+I_2}{\Delta \Phi}
\end{equation}
with $\Phi_i=i\Delta\Phi$ $(i=0,1,2)$.
For higher bias $\Phi_i$ $(i>1)$,
\begin{equation}
G_i=\frac12\left[\frac{I_{i+1}-I_i}{\Phi_{i+1}-\Phi_i}
+\frac{I_{i-1}-I_i}{\Phi_{i-1}-\Phi_i}\right].
\end{equation}

\subsection{Comparison to other methods}

\begin{figure}[bt]
\rotatebox{0}{\resizebox{3.2in}{!}{\includegraphics{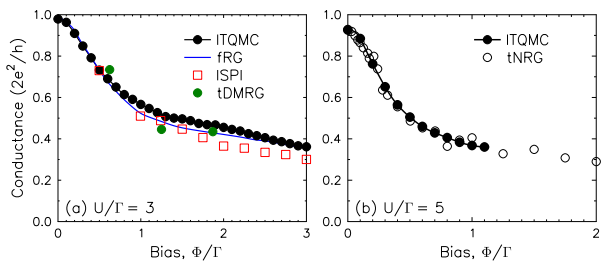}}}
\caption{
(a) Comparison of differential conductance in weakly interacting limit at
$U/\Gamma=3$ from the
imaginary-time QMC (ITQMC) at $\beta=25$, functional renormalization group
(fRG~\cite{jacobs,schmidt,eckel}, data taken from \cite{eckel}), iterative 
summation of path integral (ISPI~\cite{weiss,eckel}), 
time-dependent density matrix renormalization group
(tDMRG~\cite{feiguin}). (b) Conductance in the intermediate coupling
regime at $U/\Gamma=5$. Curves are from ITQMC ($\beta=24$) and the
time-dependent numerical renormalization group (tNRG~\cite{anders} at
$\beta=25$).
}
\label{fig:compare}\end{figure}

To demonstrate the validity of the imaginary-time QMC (ITQMC), the
differential conductance ($G=dI/dV$) is compared to other methods where
data is available.
Comparison to other methods of fRG~\cite{jacobs,schmidt,eckel},
ISPI~\cite{weiss,eckel}, tDMRG~\cite{feiguin} is shown for the weakly
interacting limit in Fig.~\ref{fig:compare}(a). In such limit the
spectral ansatz, Eq.~(\ref{eq:ansatz}), becomes an exact representation
of the self-energy at all bias and the resulting
conductance is in good agreement with other methods. Even in the intermediate
coupling limit $U/\Gamma=5$ in (b), the comparison to the time-dependent
numerical renormalization group (tNRG~\cite{anders}) results is very
good. Based on this concrete comparison and numerical efficiency of QMC
for tackling complex QD models in strongly correlated regime, this
imaginary-time method provides an efficient tool in nonequilibrium
transport theory.

\section{Results and discussions}
\label{sec:results}

\begin{figure}[bt]
\rotatebox{0}{\resizebox{3.2in}{!}{\includegraphics{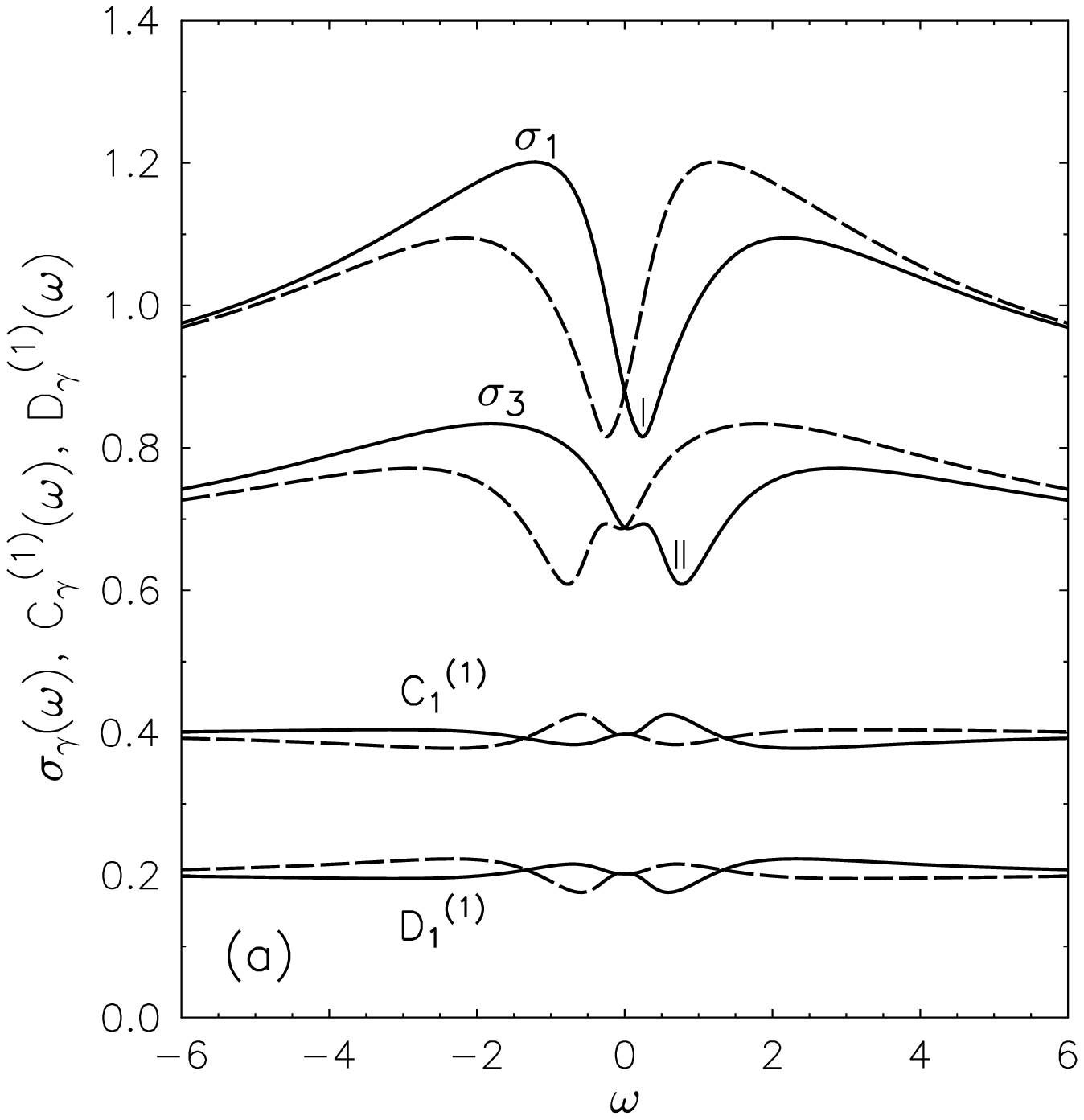}}}
\rotatebox{0}{\resizebox{3.0in}{!}{\includegraphics{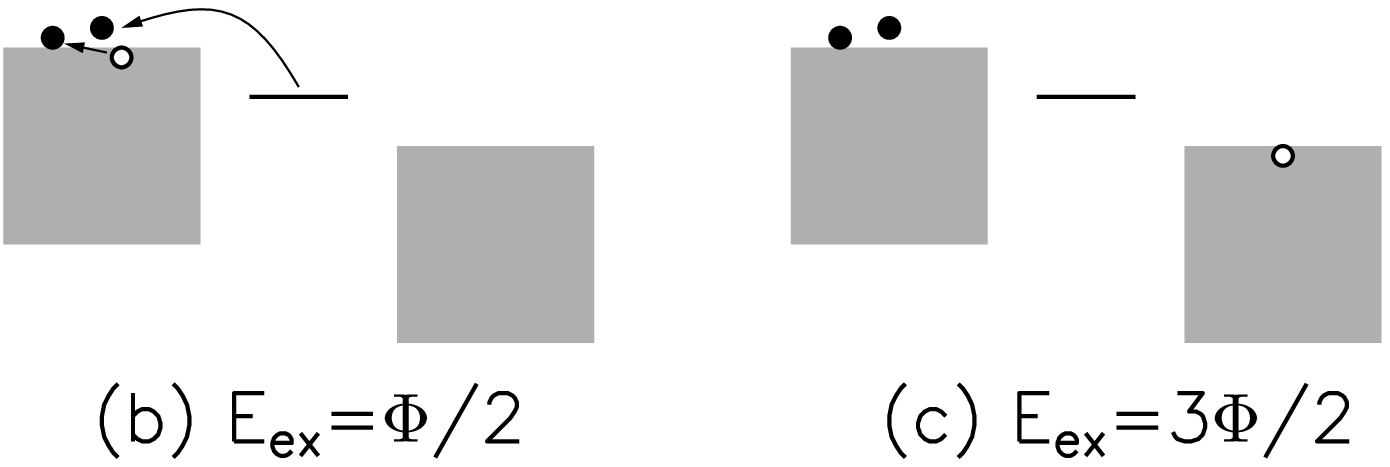}}}
\caption{
(a) Spectral functions for the ansatz
Eqs.~(\ref{eq:spec1}-\ref{eq:spec2}) at $U=10$, $\beta=36$ and
$\Phi=0.5$. Dashed lines are for negative 
$\gamma$'s, \textit{eg.} $\sigma_{-1}(\omega)$, 
$\sigma_{-3}(\omega)$, etc. As expected, $\sigma_\gamma(\epsilon)
=\sigma_{-\gamma}(-\epsilon)$ is satisfied in the particle-hole 
symmetric limit. Short vertical lines indicate $\omega=\Phi/2$ (single)
and $\omega=3\Phi/2$ (double line). (b) Intra-lead single-particle excitations
dressed by particle-hole excitations.
They contribute to $\sigma_{1}(\omega)$ at excitation energy of $\Phi/2$.
(c) Inter-lead excitations for $\sigma_{3}(\omega)$ at excitation energy
of $3\Phi/2$.
}
\label{fig:fit}\end{figure}

\begin{figure}[bt]
\rotatebox{0}{\resizebox{3.2in}{!}{\includegraphics{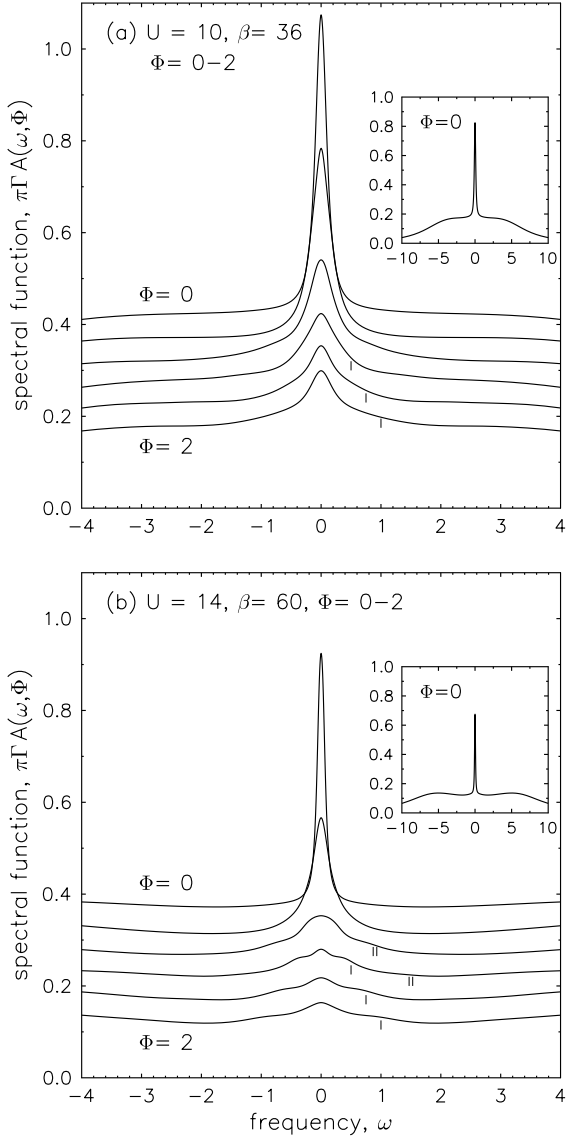}}}
\caption{(a) Spectral functions of one-particle Green's function
at $U=10$, $\beta=36$ and bias voltage
$\Phi=0,0.2,0.6,1.0,1.6,2.0$. The Kondo resonance becomes quenched
quickly as the bias $\Phi$ is applied. The spectral functions are
shifted by 0.05 for clarity.
Short vertical lines mark spectral features at
$\omega=\Phi/2$. 
Inset: spectral function for the whole frequency range.
(b) Spectral functions at $U=14$, $\beta=48$. Double vertical lines
denote the spectral contributions at $\omega=3\Phi/2$. At larger $U$ the
Kondo peaks get quenched faster to lower intensity.
}
\label{fig:U10T36}\end{figure}

The spectral functions of the self-energy from the ansatz
Eqs.~(\ref{eq:spec1}-\ref{eq:spec2}) are shown in Fig.~\ref{fig:fit} and
they demonstrate the nature of quasi-particles dressed with particle-hole
pairs. The parameters are $U=10$, $\beta=36$ and $\Phi=0.5$.
In addition to the Kondo resonance at zero frequency, there are
structures in the spectral functions at $\omega=\Phi/2$ for $\gamma=1$,
$\omega=3\Phi/2$ for $\gamma=3$, etc. For $\gamma=1$, the sum of
reservoir indices of a dressed electron has
$\sum_i\frac{\alpha_i}{2}=\frac12$ and this leads to a resonant structure at
the electron frequency measured from the combined chemical potential
$\frac12\Phi$. Similarly, there is a resonant structure at
$\omega=\frac32\Phi$ for $\gamma=3$. These effects of cross-lead
particle-hole excitations appear as shoulders in the spectral function
$A(\omega)$ in Fig.~\ref{fig:U10T36}. The magnitude of the Pad\'e
terms $C^{(1)}_\gamma(\omega)$ and $D^{(1)}_\gamma(\omega)$ is much
smaller than one, and this suggests that the higher-order Pad\'e
approximants will not significantly change the data presented here.
We also note that the symmetry relations Eq.~(\ref{eq:sym}) 
have been numerically verified.

Spectral functions in the strongly correlated limit $U=10$ and $14$ are shown in
Fig.~\ref{fig:U10T36}. In (a), the Kondo resonance develops
sharply on top of the incoherent charge excitations with the Hubbard
peaks at $\omega\sim\pm U/2$. The inset shows the spectral function for
the whole frequency range. 
The bias values are $\Phi=0,0.2,0.6,1.0,1.5$ and $2.0$ (top to bottom
curves). For clarity the curves are
shifted vertically by 0.05. The Kondo resonance is strongly quenched as
the bias is applied. There appear spectral shoulders at $\omega=\Phi/2$
(single vertical lines) and at $\omega=3\Phi/2$ (double vertical lines).
Their strength is considerably weaker than reported in
other works~\cite{ueda,anders}.

The Kondo temperature is estimated from the half-width-at-half-maximum
(HWHM) of the Kondo peak at
zero bias $\Phi=0$. Due to the incoherent charge background at 
$\pi\Gamma A(\omega)\approx 0.2$, we
read off the HWHM at $\pi\Gamma A(\omega_K)=0.6$ and
$\omega_K=0.075$. The Kondo temperature from the renormalization
group (RG) theory in the strong coupling regime~\cite{hewson,oguri} has
\begin{equation}
T_K^{RG}=\sqrt{\frac{U\Gamma}{2}}\exp\left(
-\frac{\pi U}{8\Gamma}+\frac{\pi\Gamma}{2U}
\right),
\end{equation}
with the HWHM $\omega^{RG}_K$ at
\begin{equation}
\omega^{RG}_K=\frac{4}{\pi}T_K^{RG}=0.066,
\end{equation}
in a reasonable agreement with our numerical estimate.

\begin{figure}[bt]
\rotatebox{0}{\resizebox{3.2in}{!}{\includegraphics{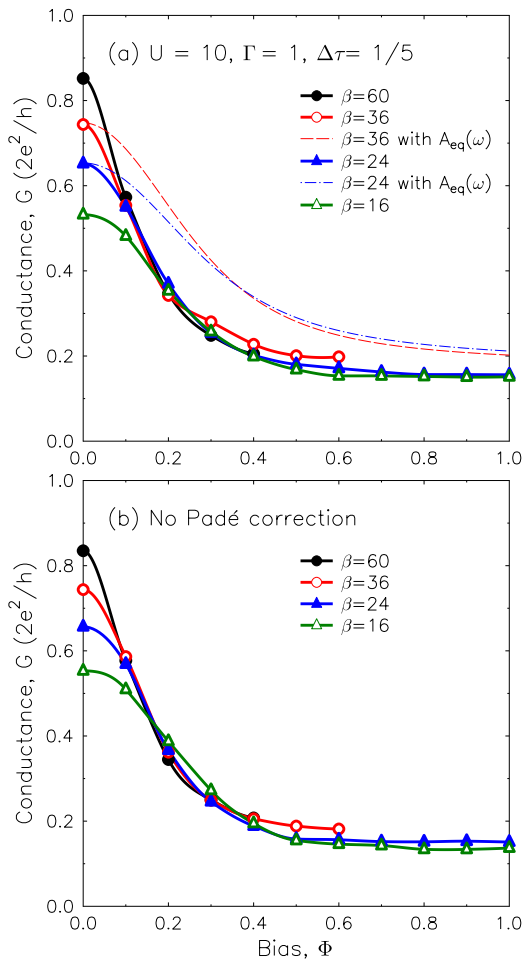}}}
\caption{(Color online) Conductance for $U=10$ at inverse temperatures
$\beta=16,24,36$ and $60$. (a) Conductance curves calculated using equilibrium spectral
function are shown with dashed lines. The narrower zero-bias
anomaly width (about $50-60$ \% from the equilibrium Kondo scale)
indicates reduction of the Kondo effect at finite bias. (b) Conductance
without the Pad\'e correction. The difference is insignificant.
}
\label{fig:U10}\end{figure}

Nonlinear conductance for $U=10$ is shown in Fig.~\ref{fig:U10} (a) with Pad\'e
approximants to the first order and (b) without the Pad\'e correction.
The comparison demonstrates that the corrections are insignificant, at
least in the particle-hole symmetric Anderson model.
Since the current is an integral
of the spectral function, details in the spectral weight shift 
tend to be insensitive to different approximations of analytic
continuation. The broken lines in (a) are derived from
Eq.~(\ref{eq:meir}) with the equilibrium spectral function
$A_{eq}(\omega)$ calculated at $\Phi=0$. Therefore, the reduced ZBA
width (about $50-60$ \%) in the full nonequilibrium calculations are due to
the destruction of the Kondo resonance at finite bias.

For a comparison to experiments and other theories, we estimate first the
temperature $T_{1/2}$ at zero bias at which the linear conductance
becomes the half conductance quantum, 
$G(T_{1/2},\Phi=0)=\frac12 G_0$.
$T_{1/2}\approx 1/14=0.071$ at a similar energy
scale with the above HWHM $\omega_K=0.075$ and $\omega^{RG}_K=0.066$.
Then, from Fig.~\ref{fig:U10T36}(a), we estimate the bias for half conductance 
quantum at the minimum temperature of simulation $T_{\rm min}$, 
$G(T_{\rm min},\Phi_{1/2})=\frac12 G_0$ is estimated to be
$\Phi_{1/2}\approx 0.135$ at $T_{\rm min}=1/60$.
From the phenomenological scaling form~\cite{grobis,oguri} of the conductance
in the leading order of temperature and bias is written as
\begin{equation}
\frac{G(T,\Phi)}{G_0}=1
-c_T\left(\frac{T}{T_{1/2}}\right)^2
-c_V\left(\frac{\Phi}{T_{1/2}}\right)^2+\cdots .
\end{equation}
By definition, $c_T=1/2$ and $c_V$ can be derived by solving
$G(T_{\rm min}=1/60,\Phi_{1/2})=\frac12 G_0$.
Then we have an estimate for the ratio of the coefficients $\alpha$  at
$U=10$ as
\begin{equation}
\alpha_{qmc}\equiv
\left(\frac{c_V}{c_T}\right)_{qmc}
\approx
\frac{T_{1/2}^2-T_{\rm min}^2}{(\Phi_{1/2})^2}\approx
0.26.
\end{equation}
Similar calculations have been repeated for $U=12$ and $14$ and the
results are summarized in TABLE~\ref{tab:alpha}. Due to the small
Kondo temperatures at large $U$,
the maximum linear conductance at zero bias
reached short of the conductance quantum at $G/G_0=0.73$ for $U=12,\beta=60$
and $G/G_0=0.63$ for $U=14,\beta=60$.
The QMC estimates for $\alpha_{qmc}$ are about $\alpha_{qmc}\sim 0.2$.
We note that the incoherent spectral background in the Anderson model
is at $0.1-0.2$ in Fig.~\ref{fig:U10T36} with the
conductance also approaching $0.1-0.2G_0$  at high bias in Fig.~\ref{fig:U10}, as
opposed to theoretical predictions from the Kondo model or renormalized 
resonant level model.
Therefore the above estimates of $T_{1/2}$ and $\Phi_{1/2}$, hence
$\alpha_{qmc}$, should be taken with some caution when compared to other
theoretical models. We also note that the estimates of
$\alpha_{qmc}$ have been derived from finite values of $T_{1/2}$ and
$\Phi_{1/2}$, instead of taking the limit $T,\Phi\to
0$~\cite{grobis,scott}.

For large $U$, the QMC calculations tend to produce overestimated Kondo
resonance HWHM, $\omega_K$, compared to $\omega^{RG}_K$. It seems that a
factor for the discrepancy is due to the discretization error despite
much improved algorithm. Part of the problem could be from the analytic
continuation where very sharp spectral peaks are fit with overestimated
width. However, it is not clear at the moment, given the above values
for $\alpha_{qmc}$,
how such discrepancy affects the scaling behaviors.

\begin{table}[h]
\begin{ruledtabular}
\begin{tabular}{cccc} 
 $U$ & $\beta_{1/2}$ & $\Phi_{1/2}/\beta_{\rm min}$ & $\alpha_{qmc}$\\ \hline
$10$ & $14$ & $0.135/60$ & $0.26$\\ 
$12$ & $21$ & $0.091/60$ & $0.24$ \\
$14$ & $36$ & $0.048/60$ & $0.21$
\end{tabular}
\end{ruledtabular}
\caption{Inverse temperature $\beta_{1/2}=1/T_{1/2}$ for
$G(T_{1/2},\Phi=0)=\frac12 G_0$, bias $\Phi_{1/2}$ for
$G(T_{\rm min},\Phi_{1/2})=\frac12 G_0$ at the minimum temperature of
$T_{\rm min}=1/\beta_{\rm min}$, and
the scaling coefficients $\alpha_{qmc}$ derived for $U=10,12,14$.
}
\label{tab:alpha}
\end{table}

In a non-interacting resonant model with a rigid spectral function
independent of temperature and bias, $\alpha_0=0.25$ can be easily
obtained. Small $\alpha$ values can be interpreted as strong temperature
dephasing of the Kondo resonance compared to that from bias voltage.
Our ratios $\alpha_{qmc}$ have larger values than the perturbative
estimate~\cite{oguri} $\alpha_{pert}=0.15$ from the effective Fermi
liquid expansion. $\alpha_{qmc}$ falls within
the experimental estimates which vary over a range of values,
$\alpha_{exp}=0.05$~\cite{scott}, $0.10$~\cite{grobis}, $0.25$~\cite{vanderwiel}.
We note that the conductance peak dies away at bias much smaller than 
where the spectral shoulders appear in Fig.~\ref{fig:U10T36}. Therefore, such
fine structures should not affect the scaling behavior. In general, the
conductance results are much more robust than spectral function
calculations.

\section{Conclusions}

Nonequilibrium imaginary-time theory has been formulated by introducing
complex chemical potentials via the Matsubara voltage. It has been
shown that the imaginary-time Green's functions upon analytic continuation
are equivalent to the real-time Green's functions. For numerical
analytic continuations we have given detailed
discussions on the analytic structure of nonequilibrium spectral
functions and generalized spectral representation in the strongly
interacting regime.
This formalism has an advantage of having familiar mathematical
structure as in equilibrium theory and can be readily adopted for
established equilibrium computational tools such as quantum Monte Carlo
(QMC) method.

Application of the Hirsch-Fye QMC method to nonequilibrium has produced
the nonlinear conductance physics where the Kondo resonance is strongly
reduced by the external bias. The conductance peak
has a reduced width from the prediction of equilibrium
calculations. By correcting discretization errors in QMC, reliable
conductance has been obtained as a function of temperature and bias.
Using a scaling form of the conductance, we obtained coefficients to the
leading temperature and bias dependent terms and their ratio $\alpha_{qmc}\sim
0.2$, larger than the perturbative prediction
$\alpha_{pert}=0.15$ but within the experimental values. This suggests 
that nonperturbative effects lead to more rapid quenching of Kondo resonance 
at finite bias.

This work shows that the imaginary-time theory provides an effective
computational tool, along with other numerical methods, in the
fast-evolving field of nonequilibrium quantum many-body theory. This
method has been applied to complex molecular quantum dot systems~\cite{prb10} 
and can be readily extended to bulk nonequilibrium using the dynamical
mean-field theory, and quantum dot systems of complex geometry. 

\section{Acknowledgements}
The author thanks Ryan Heary, Karyn Le Hur, Frithjof Anders, Akira
Oguri, Reinhold Egger and David Goldhaber-Gordon
for helpful discussions. Special thanks go to Andreas Dirks and Thomas
Pruschke who pointed out the discretization problems in QMC and made
their continuous-time QMC calculations available.
Author is grateful for financial support from the National Science
Foundation with the grant numbers
DMR-0426826, DMR-0907150 and computing resources at CCR of SUNY
Buffalo.

\begin{widetext}

\appendix 
\section{Vertex correction and branch cuts for $z_\varphi=i\varphi_m-\Phi$
variable}

\begin{figure}[bt]
\rotatebox{0}{\resizebox{3.2in}{!}{\includegraphics{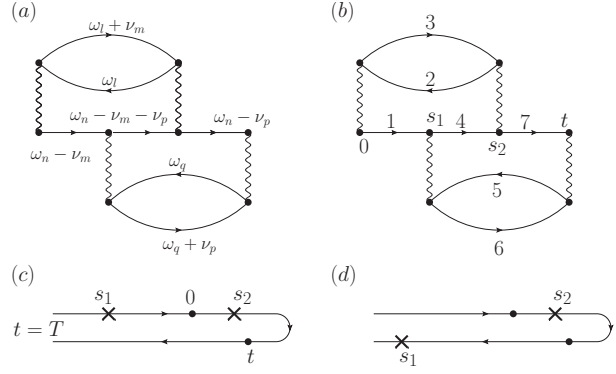}}}
\caption{(a) Fourth order vertex correction to the self-energy with the
imaginary-frequency labels. (b) The same diagram in the real-time
theory. Numerical labels denote scattering states, \textit{i.e.}
$1\equiv(\alpha_1,k_1,\sigma_1)$. (c) Time-ordering for the greater
Green's function with the interaction times $(s_1,s_2)$ with $s_1$
extending to $t=T$ on the upper Keldysh branch. (d) The same diagram
with $s_1$ on the lower branch.
}
\label{fig:pert4}\end{figure}

To examine the general analytic structure of the spectral representation of the
imaginary-time self-energy, we go beyond the second-order contribution.
Here we discuss that the spectral representation is
expressed as
\begin{equation}
\Sigma(i\omega_n,i\varphi_m)=\sum_\gamma\int d\epsilon
\frac{\sigma_\gamma(\epsilon,i\varphi_m-\Phi)}{i\omega_n-\frac{\gamma}{2}
(i\varphi_m-\Phi)-\epsilon},
\end{equation}
and how the analytic continuation $i\varphi_m\to\Phi\pm i\eta$ is taken.
Up to the second order, the spectral function
$\sigma_\gamma(\epsilon,i\varphi_m-\Phi)$ does not have any dependence
on $i\varphi_m-\Phi$. To see how
$\sigma_\gamma(\epsilon,i\varphi_m-\Phi)$ acquires the $i\varphi_m-\Phi$
dependence in the high order perturbation, we examine the vertex
correction shown in Fig.~\ref{fig:pert4}(a). First, we express the
polarization diagram $P_0(i\nu_m)$ as
\begin{eqnarray}
P_0(i\nu_m)&=&\frac{1}{\beta}\sum_{\omega_n}G_0(i\omega_n+i\nu_m)G_0(i\omega_n)\\
&=&\frac{1}{\pi^2\beta}\sum_{\omega_n}\sum_{\alpha_1,\alpha_2}\int d\epsilon_1\int
d\epsilon_2\frac{\Gamma_{\alpha_1}\Gamma_{\alpha_2}
|g(\epsilon_1)|^2|g(\epsilon_2)|^2}{(i\omega_n+i\nu_m-\frac{\alpha_1}{2}(i\varphi_m-\Phi)-\epsilon_1)(i\omega_n-\frac{\alpha_2}{2}(i\varphi_m-\Phi)-\epsilon_2)}.
\end{eqnarray}
Here we introduce short-hand notations, $\int_i=\sum_{\alpha_i}\int
d\epsilon_i$,
$\rho_i=(\Gamma_i/\pi)|g(\epsilon_i)|^2$,
$\tilde\epsilon_i=\epsilon_i+\frac{\alpha_i}{2}(i\varphi_m-\Phi)$. Then
\begin{equation}
P_0(i\nu_m)=
\frac{1}{\beta}\sum_{\omega_n}\int_1\int_2
\frac{\rho_1\rho_2}
{(i\omega_n+i\nu_m-\tilde\epsilon_1)(i\omega_n-\tilde\epsilon_2)}
=\int_1\int_2
\frac{\rho_1\rho_2(f_2-f_1)}
{i\nu_m-\tilde\epsilon_1+\tilde\epsilon_2},
\end{equation}
which can be rewritten as
\begin{equation}
P_0(i\nu_m)=
\sum_{\gamma=0,\pm 1}\int d\epsilon
\frac{A_\gamma(\epsilon)}
{i\nu_m-\gamma(i\varphi_m-\Phi)-\epsilon},
\end{equation}
with 
\begin{equation}
A_\gamma(\epsilon)=\int_1\int_2
\rho_1\rho_2(f_2-f_1)\delta(\epsilon-\epsilon_1+\epsilon_2)\delta_{\gamma,\gamma_1-\gamma_2}.
\end{equation}
The diagram in Fig.~\ref{fig:pert4}(a) becomes
\begin{equation}
\Sigma_{(a)}(i\omega_n)=\frac{1}{\beta^2}\sum_{\nu_m,\nu_p}G_0(i\omega_n-i\nu_m)G_0(i\omega_n-i\nu_m-i\nu_p)P_0(i\nu_m)G_0(i\omega_n-i\nu_p)P_0(i\nu_p).
\label{app:sig}
\end{equation}
Summing over $i\nu_m$ gives the partial factor
\begin{equation}
\int_1\int_2\int_3 \rho_1\rho_2 A_3\left[
\frac{f_1-f_2
}{(i\nu_p+\tilde\epsilon_2-\tilde\epsilon_1)(i\omega_n-i\nu_p-\tilde\epsilon_3-\tilde\epsilon_2)}
-\frac{f_1+n_3
}{(i\omega_n-\tilde\epsilon_3-\tilde\epsilon_1)(i\omega_n-i\nu_p-\tilde\epsilon_3-\tilde\epsilon_2)}
\right],
\end{equation}
with the Bose-Einstein function $n_i=(e^{\beta\tilde\epsilon_i}-1)^{-1}
=[e^{\beta(\epsilon_i-\alpha_i\Phi)}-1]^{-1}$.
Performing the summation on $i\nu_p$ on the first term proportional to
$f_1-f_2$ with the
remaining factors in Eq.~(\ref{app:sig}), we have
\begin{eqnarray}
&&\frac{1}{\beta}\sum_{\nu_p}\frac{G_0(i\omega_n-i\nu_p)P_0(i\nu_p)}{
(i\nu_p+\tilde\epsilon_2-\tilde\epsilon_1)(i\omega_n-i\nu_p-\tilde\epsilon_3-\tilde\epsilon_2)}\\
&=&\int_4\int_5\frac{\rho_4
A_5}{\tilde\epsilon_4-\tilde\epsilon_2-\tilde\epsilon_3}\left[
\frac{n_5-n_{1-2}}{\tilde\epsilon_5+\tilde\epsilon_2-\tilde\epsilon_1}
\left(
\frac{1}{i\omega_n-\tilde\epsilon_1-\tilde\epsilon_3}
-\frac{1}{i\omega_n-\tilde\epsilon_1+\tilde\epsilon_2-\tilde\epsilon_4}
\right)\right. \label{app:deno} \nonumber \\
&&\left.
+\frac{n_5+\bar{f}_{2+3}}{(i\omega_n-\tilde\epsilon_2-\tilde\epsilon_3-\tilde\epsilon_5)(i\omega_n-\tilde\epsilon_1-\tilde\epsilon_3)}
-\frac{n_5+\bar{f}_{4}}{(i\omega_n-\tilde\epsilon_4-\tilde\epsilon_5)(i\omega_n-\tilde\epsilon_1+\tilde\epsilon_2-\tilde\epsilon_4)}
\right],
\end{eqnarray}
with
$\bar{f}_i=1-f_i=(e^{-\beta\tilde\epsilon_i}+1)^{-1}=
[e^{-\beta(\epsilon_i-\alpha_i\Phi/2)}+1]^{-1}$.
This expression can be reduced to the form
\begin{equation}
\sum_\gamma\int
d\epsilon\frac{B_\gamma(\epsilon,i\varphi_m-\Phi)}{i\omega_n-\frac{\gamma}{2}(i\varphi_m-\Phi)-\epsilon},
\label{app:spec}
\end{equation}
by repeatedly using
\begin{equation}
\frac{1}{i\omega_n-z_1}\frac{1}{i\omega_n-z_2}
=\frac{1}{z_1-z_2}\left[
\frac{1}{i\omega_n-z_1}-\frac{1}{i\omega_n-z_2}
\right].
\end{equation}
It can be shown that the above form can be deduced for other types of
high order perturbation diagrams. The form Eq.~(\ref{app:spec}) 
could have been anticipated from
the Hamiltonian Eq.~(\ref{eq:k}) where $i\varphi_m-\Phi$ serves as a
\textit{parameter} and the equilibrium imaginary-time theory has a similar
spectral representation for electron self-energy due to the causality.

However there remains an important question concerning the direction
of analytic continuation of $i\varphi_m\to\Phi$. The denominator
$(i\omega_n-\frac{\gamma}{2}(i\varphi_m-\Phi)-\epsilon)^{-1}$ in
Eq.~(\ref{app:spec}) does not
pose a problem regarding the direction $i\varphi_m\to\Phi+i0^+$ or
$i\varphi_m\to\Phi-i0^+$ due to the finite imaginary number in
$i\omega_n$. However, the factor $B_\gamma(\epsilon,i\varphi_m-\Phi)$
contains energy denominators such as
$(\tilde\epsilon_4-\tilde\epsilon_2-\tilde\epsilon_3)^{-1}$ in
Eq.~(\ref{app:deno}), which may result differently depending on the
direction of the continuation $i\varphi_m\to\Phi\pm i0^+$.

To resolve this issue, we examine how such energy denominators behave in
the Keldysh real-time theory. We consider the same fourth order diagram
as shown in Fig.~\ref{fig:pert4}(b). For a specific time ordering of
Fig.~\ref{fig:pert4}(c) for $t>0$, its partial contribution to the
self-energy $\Sigma^>(t)$ can be expressed as
\begin{eqnarray}
&&\int_T^0ds_1\int_0^t ds_2
\langle d(t)d^\dagger(s_2)\rangle
\langle d(s_2)d^\dagger(s_1)\rangle
\langle d^\dagger(0)d(s_1)\rangle
\langle d(s_2)d^\dagger(0)\rangle
\langle d^\dagger(s_2)d(0)\rangle
\langle d^\dagger(t)d(s_1)\rangle
\langle d(t)d^\dagger(s_1)\rangle\\
&=&
\int_T^0ds_1\int_0^t ds_2\left[\prod_{i=1,7}\int_i\rho_i\right]
f_1 f_2 \bar{f}_3\bar{f}_4 f_5 \bar{f}_6\bar{f}_7
e^{-i\epsilon_1 s_1+i\epsilon_4(s_1-s_2)+i\epsilon_7(s_2-t)
-i\epsilon_3 s_2+i\epsilon_2 s_2 +i(\epsilon_5-\epsilon_6)(t-s_1)},
\end{eqnarray}
with the continuum labels defined in Fig.~\ref{fig:pert4}(c). Here we
take the limit $T\to -\infty$ as prescribed by Gell-mann and
Goldberg~\cite{gellmann} by taking the $T$-integral
$\eta\int_{-\infty}^0dT\,e^{\eta T}$. Also, performing the integrals on
$s_1$ and $s_2$ we get
\begin{equation}
\left[\prod_{i=1,7}\int_i\rho_i\right]
f_1 f_2 \bar{f}_3\bar{f}_4 f_5 \bar{f}_6\bar{f}_7
\frac{e^{-it(\epsilon_3-\epsilon_2+\epsilon_4-\epsilon_5+\epsilon_6)}
-e^{-it(\epsilon_6+\epsilon_7-\epsilon_5)}}{
(\epsilon_1-\epsilon_4+\epsilon_5-\epsilon_6+i\eta)
(\epsilon_3-\epsilon_2+\epsilon_4-\epsilon_7)}.
\end{equation}
Here, the bias dependence is only in the statistical factor
$f_1 f_2\cdots \bar{f}_7$.
A different contribution to $\Sigma^>(t)$ is given in
Fig.~\ref{fig:pert4}(d) with $s_1$ extending to $T$ on the lower Keldysh
branch. Its contribution can be similarly calculated as
\begin{equation}
-\left[\prod_{i=1,7}\int_i\rho_i\right]
\bar{f}_1 f_2 \bar{f}_3 f_4 \bar{f}_5 f_6 \bar{f}_7
\frac{e^{-it(\epsilon_3-\epsilon_2+\epsilon_4-\epsilon_5+\epsilon_6)}
-e^{-it(\epsilon_6+\epsilon_7-\epsilon_5)}}{
(\epsilon_1-\epsilon_4+\epsilon_5-\epsilon_6+i\eta)
(\epsilon_3-\epsilon_2+\epsilon_4-\epsilon_7)},
\end{equation}
with the negative sign coming from different Wick contraction. The only
difference from the previous expression is the statistical factor.
Regarding the convergence factor $i\eta$, we are concerned with the
contribution $A$
\begin{equation}
A=\left[\prod_{i=1,4,5,6}\sum_{\alpha_i}\int d\epsilon_i
\Gamma_{\alpha_i}|g_i|^2\right]
(f_1\bar{f}_4 f_5 \bar{f}_6-\bar{f}_1f_4 \bar{f}_5
f_6)\frac{e^{-it(\epsilon_3-\epsilon_2+\epsilon_4-\epsilon_5+\epsilon_6)}
-e^{-it(\epsilon_6+\epsilon_7-\epsilon_5)}}{
\epsilon_3-\epsilon_2+\epsilon_4-\epsilon_7}
\delta(\epsilon_1-\epsilon_4+\epsilon_5-\epsilon_6).
\label{app:A}
\end{equation}
Within the constraint given by the $\delta$-function,
\begin{equation}
f_1\bar{f}_4 f_5 \bar{f}_6-\bar{f}_1f_4 \bar{f}_5 f_6
=f_1 f_4 f_5 f_6\, e^{\beta(\epsilon_4+\epsilon_6)}
\left[e^{-\beta(\alpha_4+\alpha_6)\Phi}
-e^{-\beta(\alpha_1+\alpha_5)\Phi}\right].
\end{equation}
In equilibrium $\Phi=0$, $A=0$ and the energy integral becomes principal
value integral and the presence of $i\eta$ becomes irrelevant. The same
holds in nonequilibrium for single quantum-dot systems by applying the
same argument in Section~\ref{sec:imag} and Fig.~\ref{fig:subtle}.
States $(15)$ play the role of incoming state $|n\rangle$ and $(46)$ the
outgoing state $|m\rangle$ in Fig.~\ref{fig:subtle}(a). For example, for
$(\alpha_1\alpha_5)=(LR)$ and $(\alpha_4\alpha_6)=(LL)$ there exists a
permutation of reservoir labels in the $\alpha$-summation of
Eq.~(\ref{app:A}),
$(\alpha_1\alpha_5)=(LL)$ and
$(\alpha_4\alpha_6)=(LR)$ without permuting the energy variables
$(\epsilon_1,\epsilon_4,\epsilon_5,\epsilon_6)$ and changing the factor 
$\Gamma_1 \Gamma_4\Gamma_5\Gamma_6$. Therefore the expression $A$
becomes zero for nonequilibrium and the integrals of energy poles
around the real axis due to $i\eta$ can be replaced by principal
integrals. Finally the analytic continuation of $i\varphi_m\to\Phi$
can be taken as
\begin{equation}
(i\varphi_m\to\Phi)=\frac12\left[
(i\varphi_m\to\Phi+i0^+)
+(i\varphi_m\to\Phi-i0^+)
\right],
\end{equation}
and the subtlety of the analytic continuation $(i\varphi_m\to\Phi)$ is
resolved.

We write the total imaginary-time self-energy as
\begin{equation}
\Sigma(i\omega_n,i\varphi_m)=
\sum_\gamma\int
d\epsilon\frac{\sigma_\gamma(\epsilon)Q_\gamma(\epsilon,i\varphi_m-\Phi)}{
i\omega_n-\frac{\gamma}{2}(i\varphi_m-\Phi)-\epsilon},
\label{app:spec2}
\end{equation}
with the function $Q$ expressed as a Pad\'e quotient
\begin{equation}
Q_\gamma(\epsilon,z)
=\frac{1+C_\gamma^{(1)}(\epsilon)z+C_\gamma^{(2)}(\epsilon)z^2+\cdots}{
1+D_\gamma^{(1)}(\epsilon)z+D_\gamma^{(2)}(\epsilon)z^2+\cdots}.
\end{equation}

In particle-hole asymmetric limit, one also needs to consider the
\textit{constant} term in addition to Eq.~(\ref{app:spec2}).
\begin{equation}
\Sigma(i\omega_n,i\varphi_m)=
\Sigma^0(i\varphi_m-\Phi)+
\sum_\gamma\int
d\epsilon\frac{\sigma_\gamma(\epsilon)Q_\gamma(\epsilon,i\varphi_m-\Phi)}{
i\omega_n-\frac{\gamma}{2}(i\varphi_m-\Phi)-\epsilon}.
\label{app:spec3}
\end{equation}
$\Sigma^0(i\varphi_m-\Phi)$ is represented by another Pad\'e approximant
as
\begin{equation}
\Sigma^0(z)
=\Sigma^0\frac{1+c^{(1)}z+c^{(2)}z^2+\cdots}{
1+d^{(1)}z+d^{(2)}z^2+\cdots}.
\end{equation}
In this work, we have only considered the particle-hole symmetric limit
and the constant self-energy term $\Sigma^0(i\varphi_m-\Phi)$ has not
been included in the analytic continuation.

\end{widetext}

\end{document}